\documentclass[authoryear,nopreprintline,11pt]{elsarticle}

\usepackage{amssymb}

\usepackage{graphicx}
\usepackage{import}
\usepackage[margin=1in]{geometry}
\usepackage{booktabs} 
\usepackage{array} 
\usepackage{caption}
\usepackage{amsmath}  
\usepackage{footmisc}
\usepackage{hyperref}
\usepackage{url}      
\usepackage{xcolor}
\usepackage{bm}
\usepackage{multicol}
\usepackage{setspace}
\newcommand\fnote[1]{\captionsetup{font=footnotesize}\caption*{#1}}

\newcommand{\bftab}{\fontseries{b}\selectfont}
\newcolumntype{L}{>{\centering\arraybackslash}m{12cm}}

\begin{document}

\begin{frontmatter}



\title{Forecasting U.S. equity market volatility with attention and sentiment to the economy}



\author[mu]{Martina Halouskov{\'a}\corref{cor1}}
\author[mu,sav,fm]{Štefan Lyócsa}

\affiliation[mu]{Department of Finance, Masaryk University, Lipova 41a, 602 00 Brno, Czech Republic} 
\affiliation[sav]{Institute of Economic Research, Slovak Academy of Sciences, Sancova 56, 811 05 Bratislava, Slovakia} 
\affiliation[fm]{Faculty of Management and Business, University of Presov, Konstantinova 16, 080 01 Presov, Slovakia}
\cortext[cor1]{Corresponding author}

\fntext[fn]{This research was supported by the Czech Science Foundation (GACR), nr. 22-27075S. Computational resources were provided by the e-INFRA CZ project (ID:90254), which is supported by the Ministry of Education, Youth and Sports of the Czech Republic. We would like to thank the participants of several conferences for insightful discussions and helpful suggestions, namely, the participants of the 7th European COST Conference on Artificial Intelligence in Finance in 2022, the 28th and the 29th Annual MFS Conferences in 2022 and 2023, the 28th International Conference on Macroeconomic Analysis and International Finance in 2024, and the 26th INFER Annual Conference in 2024.}

\begin{abstract}

Macroeconomic variables are known to significantly impact equity markets, but their predictive power for price fluctuations has been underexplored due to challenges such as infrequency and variability in timing of announcements, changing market expectations, and the gradual pricing in of news. To address these concerns, we estimate the public's attention and sentiment towards ten scheduled macroeconomic variables using social media, news articles, information consumption data, and a search engine. We use standard and machine-learning methods and show that we are able to improve volatility forecasts for almost all 404 major U.S. stocks in our sample. Models that use sentiment to macroeconomic announcements consistently improve volatility forecasts across all economic sectors, with the greatest improvement of $14.99\%$ on average against the benchmark method – on days of extreme price variation. The magnitude of improvements varies with the data source used to estimate attention and sentiment, and is found within machine-learning models.

\end{abstract}



\begin{keyword}
volatility forecasting \sep investor sentiment \sep investor attention \sep macroeconomic news announcements \sep S\&P 500
\end{keyword}

\end{frontmatter}



\onehalfspacing
\section{Introduction}
\label{sec:intro}

The literature suggests that stock price changes should reflect market participants' expectations regarding the future state of the economy. This follows from the key valuation principle: the current value of an asset corresponds to discounted future payoffs to its owner. Such payoffs and discount factors should at least partly depend on the development of the economy, and therefore there is a relationship between current stock price returns and future economic growth. Empirical evidence for this relationship was provided in the early 1980s and 1990s by \citet{fama1981stock,fama1990stock,schwert1990stock}, and although it is not stable, the relationship still seems to hold according to more recent data; furthermore, it does not only hold for economic growth but also for monetary and labor market variables \citet{fromentin2022time}. Announcements of key macroeconomic fundamentals are therefore closely followed by most market participants; as such, news has the potential to change the perception of the future state of the economy, in accordance with the information arrival hypothesis. 
For example, changes in expectations about different facets of the economy change the perceived set of investment opportunities, prompting reactions from risk managers and regulators or influencing a financial manager's perception of the cost and riskiness of equity (capital).

While a large body of literature shows that stock markets react to news announcements (i.e., \citealt{flannery2002macroeconomic,ehrmann2004taking,rangel2011macroeconomic,Lucca2015,ricci2015impact,lyocsa2019central,gardner2022words}), literature that explores the role of a news announcement or perceived state of the economy in future equity price variation is limited. As noted by \citet{barbaglia2023forecasting}, individual economic indicators are noisy measures of the state of the economy that are released only monthly or quarterly (causing the ``ragged edge'' problem; see \citealt{stock2016dynamic}) and are lagging indicators in the first place. In this study, we address multiple challenges associated with forecasting equity price variation with macroeconomic variables by constructing attention and sentiment indices focused on individual key macroeconomic variables. We document considerable forecast improvements and determine not only whether attention or sentiment are useful but also what macroeconomic variables, data sources, and models are most likely to lead to the most accurate forecasts of individual U.S. stock price volatility.

In the next section, we provide an overview of the literature and the challenges associated with forecasting stock price variations. This is followed by the data section, in which we describe different data sources and attention and sentiment index creation methods. The section on methodology follows, where in addition to the standard volatility models, we describe machine learning methods, the forecasting procedures, and the forecasting evaluation framework. In the next two sections, we present our key results along with a robustness section that verifies alternative methodological choices. The final section summarizes our results and provides suggestions for future research.


\section{Literature review}

\subsection{Macroeconomic indicators and the stock market}

The perception of the future state of the economy changes continuously, with likely spikes around periods of key macroeconomic news announcements\footnote{See Figure \ref{fig:macroatt} for fluctuations in our attention measures around scheduled news announcements.}. If such news is relevant, it will likely change the valuation of the assets (e.g., \citealt{ane2000order,baker2022triggers}), which might manifest as observable price fluctuations. The effect of a news announcement on an asset's price variation might differ, depending on the stock's sensitivity to macroeconomic news announcements (e.g., \citealt{ehrmann2004taking}) or, as recently suggested by \citet{bollerslev2018volume}, depending on the level of disagreement on the interpretation of the public signal. The greater the degree of disagreement is, the less likely that increased trade will result in larger price changes; i.e., the elasticity between volume and volatility will be less than unity. 

The literature has already presented considerable evidence about market reactions to macroeconomic news announcements. For example, \citet{flannery2002macroeconomic} and \citet{rangel2011macroeconomic} provided evidence that the U.S. stock market reacts to news related to monetary policy (inflation, monetary aggregates) and the state of the real economy (e.g., unemployment and housing in \citet{flannery2002macroeconomic}). \citet{chan2018volatility} also focused on the announcement window and showed that key macroeconomic variables impact the price variation (realized and implied) of multiple assets, including the S\&P 500 market index. In the U.S. context, the role of monetary policy announcements has received considerable interest in the literature, almost unanimously showing strong ties between monetary policy announcements and reactions from U.S. equities. \citet{ehrmann2004taking} showed the heterogeneous impact of U.S. monetary policy across industries; we also observed this effect almost 20 years later, when technology stocks experienced considerable declines during the tightening of monetary policy in the early 2020s. \citet{bernanke2005explains} showed that U.S. market returns increase in response to the unanticipated interest rate cuts of the FED, and \citet{ricci2015impact} confirmed the impact of ECB announcements on the prices of large European banks. \citet{Lucca2015} explored the timing of the impact of scheduled FOMC meeting announcements on U.S. stock returns. They found that since 1980, this announcement is usually accompanied by a sharp upward preannouncement drift in the S\&P500 returns and is followed by a calm period until the next day. These studies imply that a correct forecasting setup should account for the possibility that expectations about the outcome of the given news might change not only during the announcement but also several days before and after the announcement\footnote{We should also mention the study of \citet{hussain2011simultaneous}, who showed that FED and ECB news announcements have an immediate impact on broader U.S. and European stock market indices; \citet{lyocsa2019central} also showed a strong impact of announcements by multiple monetary policy authorities (the U.S., Canada, Japan, the U.K., Europe) on the price variation of respective regional stock market indices. More broadly, \citet{megaritis2021stock} estimated macroeconomic uncertainty and successfully predicted subsequent levels of U.S. market volatility and jump-tail risk. Although their results are related to long-term market risk forecasts, they show that macroeconomic news uncertainty leads to changes in asset prices that translate into changing price variation. Previous work has revealed similar effects of different macroeconomic news announcements in the cryptocurrency market \citealt{Corbet2017, Corbet2020, AlKhazali2018TheIO, BenOmrane2023}, in the equity (\citealt{Banerjee2020}) and commodity (\citealt{Cai2020}) futures markets, and in foreign exchange markets \citealt{Boudt2019, Ayadi2020, Plihal2021}.}.

\subsection{Macroeconomic indicators and forecasting}

Several studies have considered including news announcements in their forecasting frameworks. \citet{martens2009forecasting} used several model specifications that accounted for long memory, level shifts, leverage effects, day-of-the-week and holiday effects, and macroeconomic news announcement effects, to predict the volatility of the S\&P 500. However, a comparison of models with and without news announcements (ARFI-D and ARFI-DA in Tables 4--6) revealed almost negligible improvements in terms of forecasting accuracy. \citet{rangel2011macroeconomic} reported that the event-day accuracy of volatility forecasts (see Table 8 in \citealt{rangel2011macroeconomic}) of the S\&P 500 decreased if predictive models included specific macroeconomic variables. In terms of the mean square error, improvements ranged from 2.7\% to 8.7\%, with labor market indicators being the most useful. \citet{vortelinos2015out} aimed to determine whether volatility forecasts of the U.S. equity mini futures markets lead to more accurate forecasts for days with and without one of the ten macroeconomic news announcements. Although the volatility models in \citet{vortelinos2015out} did not include macroeconomic news announcements explicitly, the results clearly indicated that volatility forecasts are more important for days on which trade balance, economic growth, price indices, or personal income statistics are reported, thus strongly supporting the idea that investors are accounting for such specific days. 
A recent study by \citet{Gupta2023} explored the predictive power of 134 macroeconomic variables grouped into 8 factors (of \cite{Ludvigson2009}) for monthly forecasts of U.S. stock market volatility. According to their results, models with macroeconomic fundamentals consistently outperform the benchmark heterogeneous autoregressive (HAR) model, and even better improvements are achieved by including survey-based measures of investor confidence and sentiment. 
Although the study focused on macroeconomic fundamentals instead of news announcements and provided evidence only for monthly forecasts, it suggests that both macroeconomic and behavioral factors play important roles in forecasting realized stock market volatility.

In this study, we utilize attention and sentiment related to key macroeconomic news, with the aim of improving predictions of price variation in the U.S. stock market. Interestingly, while the literature offers strong evidence that macroeconomic news announcements are being priced in, almost all of the empirical evidence comes from estimating pricing effects before, during, and after the news announcement, thus focusing on an event window in an in-sample fashion. For designing forecasting models, such results are useful in providing evidence about important types of news announcements (e.g., monetary policy announcements), but such models cannot be directly employed for forecasting purposes for several reasons. 

First, many specific types of macroeconomic news announcements are infrequent (e.g., quarterly GDP growth). Estimating separate parameters for each type of news (e.g., CPI, interest rate, GDP growth) will quickly increase the model complexity and likely lead to overfitted models that perform poorly in an out-of-sample setting. Second, the timing of the potential effect is unknown; i.e., we do not know how many days prior to the announcement the information about the upcoming event will be priced in. This parameter likely changes over time. Third, the importance of the news changes over time. In Figure \ref{fig:macroatt}, we visualize attention to different macroeconomic variables around corresponding scheduled news announcements. Figure \ref{fig:macroatt} shows that while attention peaks around the announcement period, it changes not only with respect to the source (e.g., Twitter, Google) and type (e.g., nonfarm payroll, FOMC meetings) of attention but also within the given macroeconomic variable. For example, since the beginning of inflation in the fall of 2021, FOMC meetings have likely been followed much more closely than before. Fourth, macroeconomic news announcements are often concentrated on certain days of the week or month and are therefore confounded with day-of-the-week effects (see \citealt{andersen1998deutsche}). Moreover, on the same day, multiple announcements can be made, and the specific timing of the announcement matters as well, e.g., whether it is before, during, or after trading hours. 

According to \citet{stock2016dynamic}, one option for addressing these issues is to construct indices that aggregate multiple macroeconomic time series into a lower-dimensional representation of the state of the economy. Indices based on announced macroeconomic data will be subject to all but the first objection made above. An alternative is to use a news-based index similar to that of \citet{baker2022triggers}, who studied drivers of large stock market price jumps. Among several interesting findings, we note that when data from major newspapers and $19$ stock market indices are used over a long period, policy-related news tends to be associated with upward price movements, especially monetary policy and government spending news. In particular, monetary policy-related news is followed by lower stock market volatility. The approach of \citet{baker2022triggers} is not suitable for forecasting purposes, as they associate news after price jumps; however, the idea of exploiting news is closely related to our approach and to those of several other studies in that we create macroeconomic news-specific attention and sentiment indices from multiple data sources (Google Trends, newspapers, the microblogging site X and Wikipedia page views). 

A separate strand of literature leverages a similar strategy, measuring attention or sentiment regarding particular topics in economic and financial news articles. Such indicators are often applied to predict macroeconomic variables instead of volatility variables. This literature includes \citet{bybee2021business}, who used news attention to explain macroeconomic dynamics and business cycles; \citet{Ellingsen2022}, who showed that news attention is particularly useful in out-of-sample forecasting of consumption; and \citet{Shapiro2022}, who estimated the impulse responses of macro variables to news sentiment shocks. \citet{barbaglia2023forecasting} also relied on six U.S. news outlets (similar to \citealt{baker2022triggers}) to construct measures of sentiment about the economy that were used as predictors of macroeconomic variables. Similarly to us, they tailored sentiment indices specifically for each variable of interest, aiming to provide a more accurate representation of the sentiment associated with the given macroeconomic variable. 

\subsection{Forecasting stock price variations with attention and sentiment}

Our approach is based on constructing attention and sentiment indices, which are theoretically founded on the concept of limited attention in \citet{kahneman1973attention} and the resulting category-learning investor behavior suggested by \citet{peng2006investor}. Since individual investors' attention is limited, investors face the difficult task of choosing from vast amounts of information, and they tend to gravitate toward "attention-grabbing" information. This led \citet{peng2006investor} to formulate category-learning behavior, and they reported that investors are likely paying more attention to macroeconomic news than to firm-specific news announcements, as it is more efficient to focus (the limited attention) on market factors that generate more uncertainty about the future investment outcomes. In accordance with these results, \citet{chen2018total} provided empirical evidence via firm-level data from the U.S. stock market that investors pay greater attention to macroeconomic news announcements than to earnings-related news. They also showed that during days with earnings announcements that are concurrent with macroeconomic news announcements, excess volatility tends to be lower than on days with earning announcements only.

Several studies have explored direct measures of attention, such as the volume of Google searches, and have shown that investor attention is correlated with lagged trading volumes \citep{Preis2010, Bordino2012}, improves trading strategies \citep{Bijl2016}, predicts stock returns  (e.g., \citealt{DA2011, Joseph2011, Bijl2016}), and predicts the market volatility of various assets (e.g., \citealt{Vlastakis2012, Kita2012, Smith2012, Aouadi2013, Goddard2015, Hamid2015, Dimpfl2016, Urquhart2018, Kim2019, Wang2018, Shen2019, audrino2020impact, Liu2022, Fiszeder2023}). Previous research has also demonstrated that different types of investor sentiment might be useful in predicting stock market volatility (e.g., \citealt{Antweiler2004, Zhang2011, Behrendt2018, audrino2020impact}, among many others). 

Another strand of literature shows that attention and sentiment related to unexpected or unprecedented events often lead to price fluctuations. 
For example, evidence was provided by \citet{maghyereh2022global} regarding the global financial crisis, by \citet{guidolin2021media} with respect to Brexit, and by \citet{lyocsa2020fear} and \citet{maghyereh2022global} regarding how attention and sentiment, respectively, 
related to the COVID-19 pandemic drove asset returns. \citet{Chen2020} further showed that Google search intensities related to coronavirus can also explain negative Bitcoin returns and high trading volume. Recently, \citet{halouskova2022role} and \citet{lyocsa2022russia} showed how attention to the Russian invasion of Ukraine in early 2022 was related to extreme asset price variations (equity and FX markets). However, all these studies are in-sample, as after the crisis event, we know what to search for; e.g., during the COVID-19 pandemic, we knew that people were searching for ``face masks'', ``travel restrictions'' and ``hand sanitizer'', but this was not relevant for markets before 2020. In this way, regular and scheduled macroeconomic news announcements offer a unique natural setting, as we know what to look for.

Among these ideas, the studies of \citet{audrino2020impact,bertelsen2021stock,AUDRINO2024103293,barbaglia2023forecasting} are most relevant. Specifically, \citet{audrino2020impact} aimed to improve volatility forecasts of U.S. equities, and in doing so, they combined multiple datasets to create attention and sentiment measures. \citet{audrino2020impact} showed strong evidence that attention and sentiment related to general interest in the stock market are important drivers of firm-level price variation, which prompted us to include such variables within our set of controls, as we are specifically interested in attention and sentiment related to terms associated with macroeconomic variables. \citet{AUDRINO2024103293} also constructed a news sentiment index from sentiment scores provided by the RavenPack News Analytics (RPNA) database. Like \citet{barbaglia2023forecasting}, they constructed indices semantically related to certain economic concepts. Their results indicate that considering news sentiment in articles on interest rates, inflation, and the labor market also improves predictions of interest rates of various maturities. We also create ten attention and sentiment indices related to specific macroeconomic variables from multiple data sources, but we study their impact on stock price variation. Finally, \citet{bertelsen2021stock} designed a smooth-transition two-regime realized GARCH model where regimes are driven by the level (or sentiment) of global (or U.S.-specific) macroeconomic news-related feeds. Their results indicate that volatility in the S\&P 500 market index is driven by the arrival of the news, regardless of whether it is positive or negative, but out-of-sample evidence is not available. 

This paper also complements the study of \citet{Bodilsen2025}. 
They show that sentiment to public news articles related to the "US economy" retrieved from proprietary macro news analytics improves volatility forecasts.
Specifically, while \citet{Bodilsen2025} derive one general sentiment score from news coverage, which is useful for forecasting individual US stocks volatility, 
we utilize both attention and sentiment from multiple sources, including social media and search engines (following \citealt{audrino2020impact}), to measure investor attitude directly related to ten specific macroeconomic announcements. Our machine learning approach allows us to identify which source of attention/sentiment to which topic impacts future volatility on a particular day. Additionally, our study confirms the impact of macroeconomic news i)  on a larger set of stocks across all S\&P500 sectors and ii) on more recent data that includes volatile market periods such as Brexit, the 2016-2020 Trump administration, and the COVID-19 pandemic, and therefore our study confirms, complements and enhances the recent results of \citet{Bodilsen2025}.

We contribute to the literature by providing multiple sets of findings that are relevant for understanding the role of macroeconomic variables in asset price fluctuations. First, we create attention and sentiment indices individually tailored to each of the ten macroeconomic variables that we cover. In our study, we show that such tailored indices are useful for forecasting purposes. With respect to the abovementioned challenges, our approach is not directly affected by the frequency of news announcements, as interest in macroeconomic variables varies over time and likely peaks around announcement days. The timing of the effect is also of no concern, as interest is estimated continuously over the full sample period, not just around news announcements. Moreover, if a specific macroeconomic variable has increased in importance, it will likely manifest in increased attention as well. Second, our work differs in scope from previous work, as our sample covers $404$ major U.S. stocks and multiple data sources and data-driven forecasting methods, which allows us to determine the effect of considering attention or sentiment, macroeconomic news sources, data sources, and industries as well as which models improve volatility forecasts the most. Third, in accordance with previous studies, we confirm that attention and sentiment toward stock markets in general are useful predictors, but attention and sentiment toward specific macroeconomic variables further improve volatility forecast accuracy. Specifically, the attention and sentiment related to FOMC meetings show the highest importance, followed by labor market indicators. Fourth, we find that analyst coverage, Google trends, news articles, and Twitter have merit in terms of prediction, whereas Wikipedia searches show little predictive utility.

\section{Data}
\label{sec:data}


\subsection{Stock market data}
Our analysis is carried out for individual constituents of the S\&P $500$ index over the period from March 10$^ {th}$, 2010, to February 24$^ {th}$, 2021, where the sample period corresponds to available data\footnote{Twitter data restrict our sample period.}. We also required each stock to have at least 2,750 trading day observations to ensure a sufficiently long time series for forecasting purposes. This led to our sample of $404$ stocks. Prices were retrieved from the FirstRade Data provider at a 1-minute calendar sampling frequency.

\subsection{Web search queries}

For more than a decade (see \cite{askitas2009google}), one of the most prevalent sources of a proxy of attention for the general public has been the Google Search Volume Index (SVI). Google provides worldwide search query data via the Google Trends platform, covering historical daily data since 2004. The data provided by Google have the form of a volume ratio on a scale of 0--100, where a value of 100 represents the highest daily number of searches for a specific term over a considered period for a given region and language. We chose to limit the search to queries made in English coming from the United States, as we aimed to capture interest in the U.S., which would be our proxy for investor trading on the U.S. stock market. Data can be downloaded at a daily frequency only if the user selects a period not exceeding 270 consecutive days; these so-called batches are always standardized between the integers 0 and 100 and cannot be naively knitted together to form a single time series (e.g., \citet{bleher2022knitting}). Since our study period is considerably longer, we downloaded data via a sliding-window technique with a window size of 270 days and an overlapping period of at least 10 days, which is used to rescale and merge individual batches into a single time series.

\begin{table}[t]

\centering
\setlength{\tabcolsep}{2.5pt} 
\small

\caption{Macroeconomic news announcements}
\label{Tbl_event}

\begin{tabular}{lllll}
\toprule
Macroeconomic indicator/event     & Variable name & N  & Rating & Ticker         \\
\midrule
Change in Nonfarm Payrolls        & NFP           & 131 & 99.21   & NFP TCH Index  \\
Initial Jobless Claims            & IJC           & 572 & 98.43   & INJCJC Index   \\
FOMC Rate Decision                & FOMC          & 88 & 97.64    & FDTR Index     \\
GDP Annualized QoQ                & GDP           & 130 & 96.85   & GDP CQOQ Index \\
CPI MoM                           & CPI           & 132 & 96.06   & CPI CHNG Index \\
ISM Manufacturing                 & ISM           & 131 & 95.28   & NAPMPMI Index  \\
U. of Mich. Sentiment             & SENT          & 263 & 94.49   & CONSSENT Index \\
Conf. Board Consumer   Confidence & CBCC          & 132 & 93.70   & CONCCONF Index \\
Retail Sales Advance MoM          & RSA           & 132 & 92.91   & RSTAMOM Index  \\
Durable Goods Orders              & DGO           & 193 & 92.13   & DGNOCHNG Index \\
\bottomrule
\end{tabular}

\end{table}

More specifically, we follow the method of \citet{lyocsa2022russia}. Let $t$ denote the usual time index, and denote consecutive batches of downloaded data by $r$; then, $S_t(r)$ is the search volume index on day $t$ found in batch $r$. In our empirical setup, each batch has $270$ consecutive days, and batches have an overlap of $10$ days
\footnote{Given our selection of keywords, the acquired batches are sometimes quite sparse; i.e., the public rarely searches for a particular keyword in a specific period, and the batches, therefore, contain many zeros. In some cases, this means that calculating a rescaling constant becomes impossible. If this occurs, we gradually extend the overlapping period between samples until we encounter at least one nonzero pair of values.}. 
For illustrative purposes, assume that a batch $r=1$ has the values $50, 60, 70, 80, 100, 40, 30, 5, 25, 5$ for periods $t=1,2,..,9, 10$ and that batch $r=2$ has $100, 21, 9, 7, 14, 13, 7, 8, 6, 5$ for periods $t = 9, 10 ,11, ..., 17, 18$. The two batches overlap in periods $t = 9, 10$. The ratio within the overlapping search volume indices leads to the following rescaling constants: $S_{t=9}(b=2) / S_{t=9}(b=1) = 4, S_{t=10}(b=2) / S_{t=10}(b=1) = 4.2$; these constants might not be the same, which occurs particularly often for smaller values because of rounding. The rounding effect leads to uncertainty related to the rescaling constant, and this problem is worse for periods in which the search volume indices swing from large to small numbers. 
To mitigate this issue, we use batches with larger overlaps, and we calculate the rescaling constant as the ratio of average values through the full overlapping period. 

The financial literature has adapted quickly to the use of Google trends data (e.g., \citealt{kristoufek2013bitcoin,preis2013quantifying}) to obtain investor attention in the form of SVIs. We follow this line of literature and use thematic lists of keywords. \textit{First}, we use a list of 70 keywords related to general stock market attention/sentiments
\footnote{\textbf{General Keywords:} Stock Market, Stocks, Bullish, Bearish, S\&P 500, Wall Street, financial markets, wall street journal, nasdaq, nyse, earnings, per share, quarterly report, earnings, call, price to earnings, price to book, market capitalisation, market price, financial times, VIX, market volatility, gold, price, t bill, treasury bill, treasury bond, 401(k), Asset Allocation, pension funds, trading volume, bear market, bull market, day, trading, technical analysis, dividend yield, futures contract, finance google, finance yahoo, marketwatch, hedge fund, market, index, mutual funds, economic recession, stop order, limit order, trading strategy, yield curve, option contract, stock symbol, market order, penny stocks, market bubble, financial crisis, market liquidity, The Motley Fool, Bloomberg.com, seeking, alpha, market downturn, volatile market, fidelity investments, etrade, ameritrade, Implied volatility, FTSE 100, Nikkei, Hang Seng Index, EURO Stoxx 50, Russell 2000, European Central Bank, EUR/USD, Eurodollar, Robinhood \label{fn:kwrds}}. 
We aim for the keywords to be general enough that they are known to the public at the beginning of the sample period (e.g., S\&P 500, Wall Street) or very specific and unlikely to be confused with other terms outside the stock market context (e.g., Robinhood, The Motley Fool). However, keyword choice uncertainty cannot be fully excluded; we addressed this by using many keywords, and the resulting SVIs are averaged for each day, representing the estimate of attention to stock markets for that day.

\textit{Second}, we 
compiled lists of keywords related to each macroeconomic variable given in Table \ref{Tbl_event_small}. We selected ten macroeconomic variables associated with high announcement importance according to Bloomberg's rating of importance, which are listed in Table \ref{Tbl_event}. This list closely follows macroeconomic variables that are often considered in the literature to study the effect of news announcements on stock price movements. For example, \cite{Lyocsa2020} used $34$ keywords associated with 8 distinct news groups (real economy, consumption, investment, government, net export, prices, monetary policy, forward-looking). As before, individual SVIs were averaged for each group of keywords related to a specific macroeconomic variable; thus, each macroeconomic variable had an associated time series of attention data from Google searches.



\subsection{Domain knowledge-specific interest: Wikipedia page views}

The next measure of attention is the daily volume of Wikipedia page views statistics. The Wikipedia API provides information on the number of views of each individual Wikipedia page, with coverage starting on December 10$^{th}$, 2007. Like the Google SVI measure, this indicator captures the volume of interest for a particular topic, but in this case, the topic is implicitly defined by the title of the Wikipedia page (see the full list in Table \ref{Tbl_wiki_keywords}). Furthermore, compared with Google SVI, Wikipedia page statistics are provided in exact numbers, meaning that there is no need for any additional construction procedures, such as merging and scaling.

\begin{figure}[!h]
\centering
\caption{Realized volatility of Apple Inc. and attention to the stock market}
\label{fig:att}
\includegraphics[width=\textwidth]{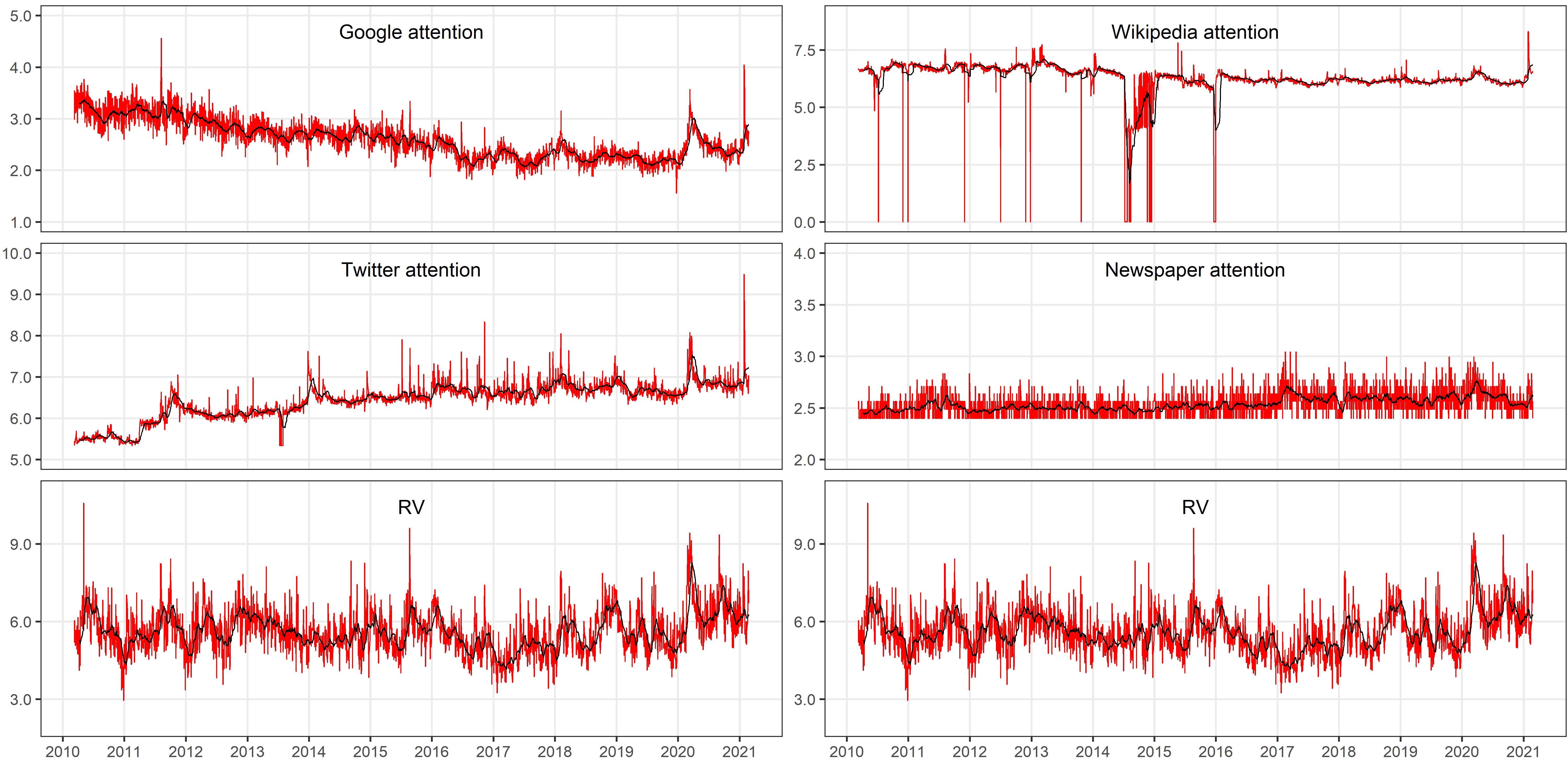}
\fnote{Notes: 
The four upper panels illustrate the time variation of four log-transformed general attention measures (in red) and their 22-day simple moving averages (in black). Clockwise, starting with the upper left panel, the figure depicts general attention to the broad stock market as indicated by Google Trends, Wikipedia, newspaper articles, and posts from Twitter.
The bottom two panels depict the log-transformed realized variance calculated from high-frequency price data of Apple Inc. (NASDAQ: AAPL; in red) along with its 22-day simple moving average (in black) throughout our whole sample period.
}
\vspace{-5pt}
\end{figure}

\subsection{Analyst attention}

As noted by \citet{baker2022triggers}, \citet{keynes1937general} suggested that investors are not interested in fundamental values but rather in the opinions of other investors about those values (see Section 5 in Chapter 12). This principle applies particularly well in forecasting, where the value itself is not very important (as it is unknown); instead, the attention and sentiment toward those variables and the information regarding on which day the corresponding news announcement is going to be made are important, as we assume that price variation might be greater before or at the day of the news announcement. We thus create daily dummy variables for each macroeconomic variable, which return a value of $1$ if the scheduled news is going to be announced before or during the trading period of the next day and $0$ otherwise. 

Bloomberg also provides information on analyst estimates related to the given scheduled macroeconomic news announcement. We use these data to construct the 'Analyst attention' variable. These variables represent how many professionals were providing estimates for the given macroeconomic variable prior to the release of the new estimates of that variable. These values are then assigned to the date on which the scheduled news is announced to capture the magnitude of analyst interest.

\subsection{Microblogging data}

Multiple microblogging platforms could be analyzed to determine investors' underlying sentiment (such as Reddit or StockTwits). Nevertheless, the leading source is the social media platform Twitter. This data source is especially suitable for several reasons. First, the Twitter API provides a rich dataset of all publicly available tweets, which can be further filtered to meet the needs of a particular research topic. Second, Twitter encompasses a broad investing community of investors, financial analysts, journalists, and others who regularly share their thoughts, views, and emotions and interact with others. Third, Twitter posts have an ideal format for sentiment analysis. Each tweet has a prescribed length of a maximum of 280 characters, meaning that the information that users try to communicate must be condensed into a few words.

Using the Academic Research Twitter API, we collected all text posts published on the website between March 2010 and February 2021 that contained specific keywords. Older data did not provide the geographical location of the account, and we did not have access to the most recent Twitter data; this determined the sample period\footnote{Since 2023, the data from Twitter have not been available for free for academic purposes.}. However, considering posts from the United States written in English and related to one of our keywords, regardless of the author of the tweet or any further restrictions, yielded a dataset of over 2 million tweets.

To capture the general interest in the stock market expressed on Twitter, we used the same list of 70 keywords as for the Google Trends data. In addition to the general interest in and sentiment toward the stock market, we are primarily interested in the attitudes of potential investors toward macroeconomic news announcements. To obtain Twitter messages that specifically discuss these ten topics, we used the topic words listed in Table \ref{Tbl_event_small}.

Then, we constructed metrics of the daily volume of messages---a general attention variable and ten macroeconomic variables---by counting all unique Twitter messages that contained some of the corresponding keywords in the text on that particular day.
Let $Z_{t,k}$ denote a tweet $k = 1, 2,..., K$ that is posted on a given day $t = 1, 2, ...$. The attention measures $T_t$ (stock market) and 
$T_{m,t}$ (macroevent specific) related to macroeconomic event $m = 1,2, ... M$, where $M=10$, are given by:

\noindent\begin{minipage}{.5\linewidth}
\begin{equation}
  T_{t} = \sum_{k \in \text{ZS}} Z_{t,k}
\end{equation}
\end{minipage}%
\begin{minipage}{.5\linewidth}
\begin{equation}
  T_{m,t} = \sum_{k \in \text{ZM}_m} Z_{t,k}
\end{equation}
\end{minipage}

where ZS and ZM denote tweets that are classified as related to either the stock market topic (ZS) or one of the $M$ macroeconomic topics. 

The content of each tweet was subsequently classified on the basis of the sentiments expressed in the text.\footnote{This process is described in more detail in Section \ref{subsubsec:sent}.} 
For each tweet $Z_{t,k}$, this determines its positive $P_{t,k}$ and negative $N_{t,k}$ sentiments, where $P \in \langle 0,1\rangle$ and $N \in \langle 0,1\rangle$. Subsequently, 22 new sentiment variables, the general positive sentiment $TWP_t$, the general negative sentiment $TWN_t$, ten positive macroeconomic variables $TWP_{m,t}$ and ten negative macroeconomic variables $TWN_{m,t}$, are defined as daily averages:

\noindent\begin{minipage}{.5\linewidth}
\begin{equation}
  TWP_t = \frac{1}{K(t)} \sum_{k \in \text{PS}} P_{t,k}
\end{equation}
\end{minipage}%
\begin{minipage}{.5\linewidth}
\begin{equation}
  TWN_t = \frac{1}{K(t)} \sum_{k \in \text{NS}} N_{t,k}
\end{equation}
\end{minipage}

\noindent\begin{minipage}{.5\linewidth}
\begin{equation}
  TWP_{m,t} = \frac{1}{K(t)} \sum_{k \in \text{PM}_m} P_{t,k}
\end{equation}
\end{minipage}%
\begin{minipage}{.5\linewidth}
\begin{equation}
  TWN_{m,t} = \frac{1}{K(t)} \sum_{k \in \text{NM}_m} N_{t,k}
\end{equation}
\end{minipage}

where PS and NS denote positive and negative sentiment values in tweets related to the general topic of the stock market, respectively, whereas $PM_m$ and $NM_m$ correspond to positive and negative sentiment in tweets within one of the $M$ macroeconomic topics. As the number of tweets $K$ is not the same every day, $K(t)$ denotes the number of tweets on a specific day.

\begin{table}[!ht]

\centering
\setlength{\tabcolsep}{2.5pt} 
\small

\caption{List of sentiment and attention measures}
\label{Tbl1_vars}

\begin{tabular}{llc}
\toprule
Variable & Description & N \\
\midrule
\multicolumn{3}{l}{\textit{Panel A: Bloomberg measures}}   \\
$B_{m,t}$ & Dummy variable of an upcoming macroeconomic news announcement        & 10 \\
$U_{m,t}$ & Number of analysts estimating the outcome of the macroeconomic news  & 10 \\
\midrule
\multicolumn{3}{l}{\textit{Panel B: General attention measures}}   \\
$ G_{t}$ & Daily average (across keywords) search volume intensity           & 1 \\
$ W_{t}$ & Daily sum of page views of a given topic                          & 1 \\
$ N_{t}$ & Daily sum of tweets with given keywords                           & 1 \\
$ T_{t}$ & Sum of articles with given keywords                               & 1 \\
\midrule
\multicolumn{3}{l}{\textit{Panel C: Individual (macro-event specific) attention measures}}   \\
$ G_{m,t}$ & Daily average (across keywords) search volume intensity         & 10 \\
$ W_{m,t}$ & Daily sum of page views of a given topic                        & 10 \\
$ N_{m,t}$ & Daily sum of tweets with given keywords                         & 10 \\
$ T_{m,t}$ & Sum of articles with given keywords                             & 10 \\
\midrule
\multicolumn{3}{l}{\textit{Panel D: General sentiment measures extracted via FinBert}}   \\
$ NPP_{t}$ & Daily average of positive sentiment in news articles            & 1 \\
$ NPN_{t}$ & Daily average of negative sentiment in news articles            & 1 \\
$ TWP_{t}$ & Daily average of positive sentiment in Twitter posts            & 1 \\
$ TWN_{t}$ & Daily average of negative sentiment in Twitter posts            & 1 \\
\midrule
\multicolumn{3}{l}{\textit{Panel E: Individual (macro-event specific) sentiment measures extracted via FinBert}}   \\
$ NPP_{m,t}$ & Daily average of positive sentiment in news articles         & 10 \\
$ NPN_{m,t}$ & Daily average of negative sentiment in news articles         & 10 \\
$ TWP_{m,t}$ & Daily average of positive sentiment in Twitter posts         & 10 \\
$ TWN_{m,t}$ & Daily average of negative sentiment in Twitter posts         & 10 \\
\bottomrule
\end{tabular}

\fnote{Notes: Notation of all attention and sentiment measures used in this study. Note that the last column includes the number of measures corresponding to each category of variables. Each measure of an individual (macro-event specific) attention or sentiment is specified as related to $mth$ macroeconomic news announcement.}

\end{table}

\subsection{Financial news data}

Using the ProQuest database, we collected news articles related to general trading activities and the $10$ selected macroeconomic variables that appeared in two major financial newspapers---The Financial Times and The Wall Street Journal---and then narrowed down the results to U.S. news only. Our selection of individual newspapers follows the approach of others (e.g., \citealt{Tetlock2007, DA2011, Goddard2015, baker2022triggers}).

We searched newspaper articles on the basis of (1) general keywords (listed in footnote \ref{fn:kwrds}) and (2) lists of keywords corresponding to the ten macroeconomic events (in Table \ref{Tbl_event_small}) to obtain both general and event-related attention and sentiment variables. Specifically, we downloaded each article mentioning the keyword or key phrase in its title, abstract, or text. This procedure resulted in 6,660 articles in total. Finally, we constructed measures of attention and sentiment, following the same procedure described in the previous section on microblogging data.

Figure \ref{fig:att} presents the study's 
attention variables 
related to the stock market in general, along with the realized volatility of Apple Inc. (NASDAQ:AAPL), which is one of the examined stocks. The figure covers the entire sample period and shows all variables in question at a daily frequency and after a log transformation. Although the attention measures have different behaviors, they all, in many instances, closely follow the upward or downward movements of realized volatility.

\subsection{Sentiment extraction}
\label{subsubsec:sent}

In this section, we describe procedures for sentiment extraction from news articles and tweets. In this study, we rely on a state-of-the-art NLP model, FinBERT \citep{Araci2019}, which is a language model designed for financial sentiment classification. FinBERT is an extension of the bidirectional encoder representations from transformers (BERT) model, a language model with a transformer architecture that was introduced in 2018 by \citet{Devlin2019} and has become one of the most utilized natural language processing (NLP) models. FinBERT was specifically pretrained on a large corpus of financial texts from the Financial PhraseBank by \citet{malo2014good}. Owing to this pretraining, FinBERT is fine-tuned for the classification of sentiment in financial texts. 

Before sentiment classification, we applied several text-cleaning and preprocessing procedures to obtain a format that can be further analyzed. However, since we chose a rather advanced sentiment model, we performed only some essential text-cleaning tasks, allowing FinBERT to do most of the work and keeping as much context as possible. We removed all website URLs, several special characters, and punctuation while keeping intraword contractions and dashes. Following the necessary preprocessing steps, the pretrained FinBERT model was applied to each cleaned text document separately, which yielded ratios of positive and negative sentiments expressed in the text, ranging from 0 to 1. Finally, we created daily sentiment scores (listed in Table \ref{Tbl1_vars}) by averaging the sentiment ratios of tweets or news articles published on a given day in a particular category of keywords. In the robustness section, we also consider alternative strategies for sentiment classification.

\section{Methodology}
\label{sec:methodology}

\subsection{Volatility estimators}
\label{sec:estimators}

We study the forecasting power of sentiment and attention indicators related to macroeconomic variables to predict the price variation of U.S. equities. To estimate intraday price variation, we use the realized variance (RV). Specifically, let $P_{t,j}$ correspond to equally spaced (calendar sampling) prices of the asset on day $t=1,2,...$ at intraday time $j=0,1,..., M$, where for a 
$5$-minute sampling frequency, $M=78$. At $j=0$, we have the initial (opening) price, and at $j=78$, we have the last (closing) price; thus, we have $N=M-1$ returns a day. The intraday continuous return is given by $R_{i,j} = ln(P_{t,j}) - ln(P_{t,j-1})$, and the intraday realized variance is given by:
\begin{equation} \label{eq:rv}
    RV_{t} = \sum_{j=1}^{N} R_{t,j}^{2}
\end{equation}
A $5$-minute sampling frequency is the most popular choice, as it is less prone to potential biases arising from microstructure noise effects (e.g., bid-ask bounce) that are observable at higher sampling frequencies. The choice of a $5$-minute sampling frequency is standard in the literature (see \citealt{liu2015does}). Given that our price series is available at $1$-minute sampling frequencies, there are $5$ possible 'grids' that we can use to estimate intraday price variation. We follow \citet{Patton2015} and take the average of the correlated but not identical intraday realized variance to increase the efficiency of the estimator. In the following, we denote this average as $RV_t^{IN}$.

The closing price rarely matches the next day's opening price, suggesting that the valuations also change outside of official trading hours, although these changes are not explicitly observed. Macroeconomic variables might also be reported before or after official trading hours. We should therefore consider the overnight price variation $RV_t^{ON}$ and predict the whole-day price variation $RV_t$, not just the intraday price variation $RV_{t}^{IN}$. We follow the approach of \citet{Hansen2005} by estimating weights for the overnight ($\omega_1$) and intraday ($\omega_2$) price variations. First, we define the overnight return as $RV^{ON}_{t,j}=\left[ln(P_{t,0}) - ln(P_{t-1,M})\right]^2$ and annualize the overnight squared return as $RV^{ON}_{t}=r^{ON,2}_{t,j}\times100^{2} \times 252$. The RV measure for a full day is then given by a weighted sum of the intraday and overnight components:
\begin{equation} \label{eq:rvHL}
    RV_{t}=\omega_{1}RV_{t}^{ON} + \omega_{2}RV_{t}^{IN}
\end{equation}

Price variation can occasionally experience large or small sudden shifts, which are often referred to as jumps. In terms of forecasting future price variation, capturing these jumps has proven to be beneficial \citep{Andersen2007, Corsi2010}. We decompose the intraday price variation into two parts, following \citet{Andersen2012}: the jump component ($JC_{t}$) and the continuous component ($CC_{t}$), which can be defined as:
\begin{equation} \label{eq:jc}
    JC_{t}=max\{0,(RV_{t}^{IN}-MedRV_{t})I[JT_t>\phi_{1-\alpha}]\}
\end{equation}
and:
\begin{equation} \label{eq:cc}
    CC_{t}=MedRV_{t}I[JT_{t}>\phi_{1-\alpha}] + RV_t^{IN}[JT_{t}\leq\phi_{1-\alpha}]
\end{equation}
where $I[.]$ is the indicator function, returning $1$ if the condition is true and $0$ otherwise. $\phi_{1-\alpha}$ is the $1-\alpha$ quantile of the standard normal distribution. $MedRV_{t}$ is the median realized jump-robust\footnote{In the presence of price staleness, realized multipowers are biased estimators of integrated volatility powers (see \citealt{Kolokolov2024}). However, our sample consists of highly liquid stocks. Additionally, the jump estimates are only one of the features in the CSR HAR model, and other models demonstrate similar forecasting performance.} volatility estimator:
\begin{equation} \label{eq:mrv}
    MedRV_{t}=\frac{\pi}{6-4\sqrt{3}+\pi}\left(\frac{N}{N-2}\right) \sum_{j=2}^{N-1} [med|R_{t,j-1}|,|R_{t,j}|,|R_{t,j+1}|]^2
\end{equation}
and $JT_{t}$ is the test statistic given by:
\begin{equation} \label{eq:jt}
    JT_{t}=\sqrt{M}\frac{(RV_t^{IN}-MedRV_t)(RV_{t}^{IN})^{-1}}{(0.96max\{1,MedRQ_t/MedRV_{t}^{2}\})^{1/2}}
\end{equation}
where the median realized quarticity $MedRQ_{t}$ is:
\begin{equation} \label{eq:mrq}
    MedRQ_{t}=\frac{3\pi N}{9\pi+72-52\sqrt{3}}\left(\frac{N}{N-2}\right) \sum_{j=2}^{N-1} [med|R_{t,j-1}|,|R_{t,j}|,|R_{t,j+1}|]^4
\end{equation}

It follows that if the jump component is not statistically significant, the continuous component is equal to the intraday price variation. We next follow the work of \cite{Patton2015}, who used the semivariances of \citet{10.1093/acprof:oso/9780199549498.003.0007} to improve volatility forecasts. The semivariances are given as:
\begin{equation} \label{eq:RS}
\begin{aligned}
    RS_{t}^{+} = \sum_{j=1}^{N} R_{t,j}^{2}I[R_{t,j>0}] \\
    RS_{t}^{-} = \sum_{j=1}^{N} R_{t,j}^{2}I[R_{t,j\leq0}]
\end{aligned}
\end{equation}

where the $I[.]$ indicator function takes a value of 1 if the condition holds and 0 otherwise. The semivariances can also be used to estimate the signed jump $SJ_t$. Specifically, given the underlying symmetrical data generating process (see \citealt{10.1093/acprof:oso/9780199549498.003.0007,Patton2015}) $RS_{t}^{+}$ and $RS_{t}^{-}$, we can eliminate the continuous part and obtain the so-called signed jump, which is used in HAR-SJ, a model proposed by \citet{Patton2015}.

\begin{equation} \label{eq:SJ}
\begin{aligned}
    SJ_{t}=RS_{t}^{+}-RS_{t}^{-},
\end{aligned}
\end{equation}

We use $N$ average returns at a $5$-minute sampling frequency (calculated by averaging across multiple sampling grids as described above) to estimate the jump component $JT_t$, continuous component $CC_t$, semivariances $RV_t^{+}, RV_t^{-}$ and signed jump $SJ_t$.

\subsection{Benchmark volatility models}
\label{sec:bench}

In this study, we employ three different benchmark models. The first model is the standard realized volatility heterogeneous autoregressive (HAR) model developed by \citet{corsi2009simple}, the structure of which approximates the heterogeneous behavior of individual market participants with three different investment horizons:
\begin{equation} \label{eq:har}
    RV_{t+1}= \beta_{0} + \beta_{1}RV_{t}^{D} + \beta_{2}RV_{t}^{W} +\beta_{3}RV_{t}^{M} + \epsilon_{t+1}
\end{equation}

where $RV_t^D, RV_t^W$ and $RV_t^M$ are nonoverlapping averages of realized variance over 1, 5 (weekly), and 22 (monthly) days.

Next is the HAR-M model, which is the standard HAR model augmented with ten macroeconomic dummy variables (see Section 2.4) that indicate whether the next price variation coincides with a given macroeconomic news announcement:
\begin{equation} \label{eq:har-m}
\begin{aligned}
    RV_{t+1}= \beta_{0} + \beta_{1}RV_{t}^{D} + \beta_{2}RV_{t}^{W} +\beta_{3}RV_{t}^{M} + 
    \sum_{m=1} \gamma_m RV_t^{D} B_{t,m}
   + \epsilon_{t+1}
\end{aligned}
\end{equation}

For the final benchmark model, we construct a CSR HAR model following \cite{lyocsa2021improving}, which is based on the complete subset regression of \citet{elliott2013complete}. The idea is to combine forecasts from multiple benchmark models into a single-point forecast, thus exploiting the potential predictive powers of different components of the price variation. The technical details are described in Section \ref{subsubsec:csr}. 
This benchmark model includes a signed jump, jump, continuous components, and negative semivariances, thus potentially encompassing the HAR-SJ and HAR-SV models of \cite{Patton2015} and the HAR-CJ model of \cite{Andersen2007, Andersen2012}. All of these components are considered as daily observations as well as weekly and monthly averages. The overall number of variables used in the complete subset regression framework was $15$.

\subsection{Competing models: Sentiment and attention volatility models}

Against the three benchmark models, we set up eight competing models: i) four models for attention only and ii) four models including positive and negative sentiments. The four models are an augmented HAR model, the adaptive least absolute shrinkage and selection operator (LASSO) (e.g., \citealt{audrino2020impact}), the complete-subset HAR model (e.g., \citealt{lyocsa2021improving}), and a random forest (e.g., \citealt{christensen2023machine}). 

\subsubsection{Augmented HAR models}
The straightforward competing model is based on the augmentation of the standard HAR model with external information from measures that represent attention to the general stock market or macroeconomic variables, news arrival and sentiment expressed by users on social media and observed by investors in financial news articles. Given the $10$ macroeconomic news announcements, multiple data sources, and attention and sentiment, we have $96$ potential explanatory variables, which would make the HAR model highly inefficient. For the HAR class of competing models, we take the average across corresponding attention or sentiment measures (see Table \ref{Tbl1_vars} for variable descriptions). Specifically, the HAR-A model is augmented with attention to the stock market only:
        \begin{equation}\label{eq:haratt}
        \begin{aligned}
            RV_{t+1}^{d} =\beta_{0} + \beta_{1}RV_{t}^{D} + \beta_{2}RV_{t}^{W} +\beta_{3}RV_{t}^{M} +\delta_{1}G_{t} + \delta_{2}W_{t} + \gamma_{3}T_{t} + \gamma_{3}N_{t} + \epsilon_{t+1}
        \end{aligned}
        \end{equation}
        
The HAR-S model includes both positive and negative sentiment toward the stock market: 
        \begin{equation}\label{eq:harsent}
        \begin{aligned}
            RV_{t+1}^{d} =\;&\beta_{0} + \beta_{1}RV_{t}^{D} + \beta_{2}RV_{t}^{W} +\beta_{3}RV_{t}^{M} \;+ \\ 
            &\gamma_{1}TWP_{t}+ \gamma_{2}NPP_{t} +\gamma_{1}TWN_{t} + \gamma_{2}NPN_{t}  + \epsilon_{t+1}
        \end{aligned}
        \end{equation}
All HAR models are estimated via weighted least squares (WLS), where the weights are optimal in the presence of an unknown structural break in the coefficients (\citealt{pesaran2013optimal}). The weights correspond to a smooth, exponentially declining function that gives higher weights to more recent observations.

Neither of these models includes attention or sentiment related specifically to macroeconomic variables/news announcements. Comparing these models against the benchmarks shows the merit of attention and sentiment toward stock markets, which was shown by \citealt{audrino2020impact} to be a relevant predictor, albeit on a smaller sample. In what follows, we present competing models that include, in addition, either attention or sentiment related to macroeconomic news announcements. We use adaptive LASSO, complete subset regression, and random forest, as these machine-learning-based modeling approaches can utilize a subset of the $96$ attention or sentiment variables.

\subsubsection{Adaptive Lasso}

We use adaptive LASSO, proposed by \citet{zou2006adaptive}, which is an extension of the popular LASSO model of \citet{tibshirani1996regression}. Adaptive LASSO is, in essence, a two-stage LASSO estimator. In the first stage, we employ the ridge estimator (\cite{hoerl1970ridge}) to find the coefficients:
\begin{equation}
\label{eq:ridge}
    (\hat{\bm{\beta}},\hat{\bm{\gamma}}^{*}) = 
    \operatorname*{arg\,min}_{\bm{\beta},\bm{\gamma}} \left\{\sum_{t=1}^{T} \left(RV_{t+1} - \bm{\beta}^{\top} \bm{RV}_t - \bm{\gamma}^{\top} \bm{X}_{t,k}
    \right)^{2} 
    + \lambda_{init}\left( \bm{\gamma}^{\top}\bm{\gamma}\right) \right\}  
\end{equation}
where $\bm{RV}_t = (1, RV_{t}^D, RV_{t}^W)^{\top}$ is a column vector with intercept and lagged volatility terms, $\bm{X}_{t,m} = (X_{t,1}, X_{t,2}, ..., X_{t,p})^{\top}$ is a column vector of additional variables of interest (e.g., monthly volatility, attention or sentiment variables), and the corresponding coefficients are $\bm{\beta}_{m} = (\beta_{0}, \beta_{1}, \beta_{2})^{\top}$ and $\bm{\gamma} = (\gamma_{1}, ...,\gamma_{p})^{\top}$. The size of the estimated coefficients is penalized by the sum of squared coefficients $\left( \bm{\gamma}^{\top}\bm{\gamma}\right)$, the $l_2 - norm$ penalty term. The coefficients move toward zero more quickly the larger the parameter $\lambda$ is, whereas $\lambda = 0$ produces classical least squares estimates. The coefficients estimated from the first stage are used in the second stage to calculate coefficient-specific penalty weights for the second step of the adaptive LASSO as follows:
\begin{equation}
\label{eq:weights}
    w_k = \frac{1}{|\hat{\gamma}_{k}^{*}|}, k=1, ... ,p
\end{equation}

Stacking the weights into a column vector $\bm{w} = (w_1,w_2,...,w_k)^{\top}$, the estimator is obtained as:
\begin{equation}
\label{eq:adalasso}
    (\hat{\bm{\beta}}^{A}, \hat{\bm{\gamma}}^{A}) = 
    \operatorname*{arg\,min}_{\bm{\beta},\bm{\gamma}} \left\{\sum_{t=1}^{T} \left( RV_{t+1} - \bm{\beta}^{\top} \bm{RV}_t - \bm{\gamma}^{\top} \bm{X}_{t}
    \right)^{2} 
    + \lambda_{adap} \left( \bm{w}^{\top}|\bm{\gamma}|\right) \right\}  
\end{equation}
where $|\bm{\gamma}|$ denotes that all the elements of vector $\bm{\gamma}$ are in absolute terms. In the estimation described above, all the variables are standardized so that the penalization has an equal effect on all the variables, irrespective of their level or variance. Moreover, we use weighted estimation, where observations receive weights in a similar way to those of the augmented HAR models (as discussed in \citealt{pesaran2013optimal}). 

We have two types of adaptive LASSO models that differ in the set of employed variables. The first uses all attention variables, and the second uses all sentiment variables.

\subsubsection{Complete Subset Regression}
\label{subsubsec:csr}

Next, we consider the complete subset regression approach developed by \citet{elliott2013complete} and recently applied by \citet{lyocsa2021improving} in a volatility forecasting context. The general idea is to combine predictions generated from all possible (subset) specifications with given numbers of potential regressors and parameters.
If the number of all regressors is $p$ and $k \leq p$ (the complexity parameter) denotes the number of variables in a model $m$, a forecast generated from model $m = 1, 2, ..., \frac{p!}{(p-k)!k!}$ is given by:
\begin{equation}
    RV_{t+1,k,m}=\bm{\beta}_{k,m}^{\top} \bm{RV}_t + \bm{\alpha}_{k,m}^{\top} \bm{X}_{t,k,m} + \epsilon_{t+1,k,m}
\end{equation}
where $\bm{RV}_t = (1, RV_{t}^D, RV_{t}^W)^{\top}$ is a column vector with intercept and lagged volatility terms, $\bm{X}_{t,k,m} = (X_{t,k,m_1}, X_{t,k,m_2}, ..., X_{t,k,m_k})^{\top}$ is a column vector of additional variables of interest (e.g., monthly volatility, attention or sentiment variables), and the corresponding coefficients are $\bm{\beta}_{k,m} = (\beta_{0,m}, \beta_{1,m}, \beta_{2,m})^{\top}$ and $\bm{\alpha}_{k,m} = (\alpha_{m,1}, ...,\alpha_{m,k})^{\top}$. We let $k$ be equal to $2$; thus, in addition to the daily and weekly volatility components, we try all the combinations of the remaining variables. For example, for $30$ potential explanatory variables, we have $435$ potential models. 

Because we use forecasts ($\widehat{RV}_{j,k,m}$) from many linear regression models $(k,m)$, we need a rule to aggregate the forecasts into one estimate. A simple approach would be to take an average. However, if we have a small subset of well-performing models, the average will not work very well. Alternatively, if many of the models are biased, the average forecast will be biased, and the bias will likely not be compensated for with increased efficiency. Moreover, some of the models may be predictably inaccurate. For these reasons, we use the discounted mean square error (DMSE) recommended by \cite{stock2004combination}, with $\delta = 0.95$ over the calibration sample of $S=500$ out-of-sample observations for all models $(k,m)$, to determine which models perform better:
\begin{equation}
    w_{t,k,m}(\delta) = \left[S^{-1} \sum_{j=t-S+1}^{t} \delta^{t-j} (RV_{j} - \widehat{RV}_{j,k,m})^2\right]^{-1}
\end{equation}

The final prediction is not a weighted average, as in a finite sample, this would be biased as well. Instead, we aim to identify a group of models with the most accurate forecasts. To do so, we use $5$-means clustering to find which models belong to a cluster of models with the lowest DMSE. The forecast in the next period $t+1$ is a 25\% trimmed mean of the forecasts generated by models that belong to the cluster of most accurate models. If the cluster has fewer than $4$ models, we take a simple average instead of a trimmed average. Note that we have two types of complete subset regression models that differ with respect to the employed set of regressors, where we use either only attention or only sentiment variables.

\subsubsection{Random Forest}

The random forest, proposed by \citet{breiman2001random}, is designed to decrease the correlation in regression trees. The random forest uses bagging, an algorithm that builds $B$ regression trees over a collection of $B$ different bootstrap training subsamples. Let $p(T_b(\bm{X}))$ denote a prediction from a tree $T_b(\bm{X})$ with a given set of regressors $\bm{X}$ and a bootstrap sample $b$. The trees are built via recursive binary splitting to minimize the mean square error. The trees are grown until a split cannot occur, as the number of observations would fall below $10$ in the subsequent leaves. This leads to low-bias but high-variance predictions. To improve the efficiency, the predictions are averaged across all trees generated from different bootstrap samples:
\begin{equation}
    f_{RF}^{B} (x) = \frac{1}{B} \sum_{b=1}^{B} p(T_{b}(x))
\end{equation}

The predictions are likely highly correlated, especially if only a minor subset of regressors is relevant. The improvement by \cite{breiman2001random} is that during each split, instead of considering all features, a random set of $z$ features is considered. By limiting each tree to finding the best predictor out of the $z$ random features, the resulting predictions will be less closely correlated.

Random forests require hyperparameter tuning. In our setting, every estimation uses $B=500$ trees; we consider $z \in (8,16,32)$ and the maximum depth of the tree $d \in (6, 12, \text{max})$, where $\text{max}$ corresponds to the maximum depth before the condition of at least $10$ observations in the terminal node is reached. This leads to 9 possible settings. Similar to adaptive LASSO, we leave the daily and weekly realized variance components out of the method by forcing the algorithm to always consider these variables in addition to the $z$ randomly selected regressors. We use a calibration sample of $500$ consecutive observations to select the one model out of nine that leads to the lowest forecast error (see the Forecast Evaluation section). As before, we have two random forest models. Either the model uses only attention variables or only sentiment variables.

\subsection{Variable transformations}

Stock price variation is known to be highly right skewed and subject to infrequent but extremely high values. Similar characteristics can be observed across almost all stocks and for other components of the price variation process. This tends to lead to a positive relationship between the observed level of price variation and the forecast error. An alternative is to use a log specification, which is discussed as one of the preferred approaches in \cite{clements2021practical}. Therefore, in all the models, we apply the logarithmic transformation\footnote{Specifically, if $x$ is the given variable and $x=0$, the transformation is $ln(x+1)$; otherwise, we use $ln(x)$.}
to all components of the price variation: the realized variance, signed jump, jump, continuous components, and semivariances. We also use the log transformation for attention and sentiment measures. This is a common practice in the literature on forecasting market volatility (e.g., \citealt{audrino2016lassoing,clements2017forecasting,audrino2020impact}). By taking the log, we transform these variables into ones with more symmetrical distributions, while the effect of the outliers on the parameter estimates is reduced considerably. All variables except for the signed jumps are nonnegative. The components of signed jumps require special treatment. Specifically, the transformed signed jumps are as follows:
\begin{equation} \label{eq:lnsj}
   \begin{cases}
       \ln(SJ_t),    & \text{if } SJ_t > 0 \\
       \ln(1+SJ_t),  & \text{if } SJ_t = 0 \\
	   -\ln(\lvert SJ_t \rvert), & \text{if } SJ_t < 0 \\
       \end{cases}
\end{equation}

To avoid confusion, we do not change the notation for the variables; however, from this point onward, we refer to the transformed price variation as volatility and the untransformed version as the price variation or variance.

\subsection{Forecast procedure and evaluation}
\label{subsec:forecast}

The forecasting procedure is based on a rolling-window forecast with an estimation window of $1000$ observations. Complete subset regression, adaptive LASSO, and random forest require a calibration sample to tune the hyperparameters. The sample is set to $500$ observations. It follows that the first forecast falls on the $1501^{th}$ day, which for most of the stocks corresponds to 24 February 2016; most of the stocks have approximately $1260$ out-of-sample observations, i.e., 5 years of data. 

As discussed in the previous section, we predict the log of the variance. To evaluate the untransformed price variance, we need to transform the predicted log variance. We use the second-order approximation as described in \cite{taylor2017realised}. 

Furthermore, the HAR model is known to occasionally produce either unreasonably large or small forecasts. If this happens, we apply a simple "insanity filter" (see, e.g., \citealt{SWANSON1997439}), which ensures that the forecasted value is no smaller and no larger than the values observed in the estimation window. The same filter is applied to all competing models.


The out-of-sample forecast accuracy is assessed via one of the two most widely used loss functions---the mean squared error (MSE) defined in Equation \ref{eq:mse}---as suggested by \cite{Patton2011}. We further explore our out-of-sample results under alternative loss functions as part of the robustness checks in Subsection \ref{subsubsec:losses}.
\begin{equation}
\label{eq:mse}
    MSE=n^{-1}\sum_{t=1}^{n} (RV_{t+1}-\widehat{RV}_{t+1})^2
\end{equation}

The main results in the text are presented under the $MSE$ loss function. A statistical comparison across all the models is run via the model confidence set (MCS) approach as described by \cite{Hansen2011}. Under the MSE, the loss differential $d_{i,j,t} = (RV_{i,t+1}-\widehat{RV}_{i,t+1})^2 - (RV_{j,t+1}-\widehat{RV}_{j,t+1})^2$ is evaluated. We are interested in a test of the null hypothesis $H_{0,\mathcal{M}}: \mu_{ij}=0 $, where $\mu_{ij} \equiv E(d_{ij,t})$, e.g., the expected value of the loss differential between models $i$ and $j$ for all $i,j \in \mathcal{M}$ at period $t$. The MCS procedure starts with a set of competing models $\mathcal{M}$, which are eliminated through a series of equivalence tests until we obtain a set of superior models $\mathcal{M^*}$ that contains the best models with a given confidence. The procedure stops as soon as we are unable to reject the null hypothesis of the equivalence of the remaining models.

\section{Results}
\label{sec:results}

\subsection{Preliminary data analysis}

Before we proceed with our main analysis, we explore the properties of the key variables of interest. Tables \ref{Tbl1}, \ref{Tbl1_macro_att} and \ref{Tbl1_macro_sent} report the key statistical characteristics of our data.

\begin{table}[!ht]

\centering
\small

\caption{Descriptive statistics of realized measures, four general attention, and four general sentiment measures.}
\label{Tbl1}
\begin{tabular}{lrrrrrrrrrrrr}
\toprule
Variable       & Mean  & S.D. & Skew. & Kurt. & Min.  & 5th   & 50th  & 95th  & Max.  & $\rho(1)$ & $\rho(5)$ & $\rho(22)$ \\
\midrule
\multicolumn{13}{l}{\textit{Panel A: Realized measures}}                                                     \\
$RV_{t}^{D}$  & 6.02  & 0.85 & 0.93  & 5.11  & 3.70  & 4.84  & 5.92  & 7.54  & 10.54 & 0.72 & 0.56 & 0.38  \\
$JC_{t}^{D}$  & 4.86  & 1.10 & -1.40 & 10.61 & 0.01  & 3.48  & 4.88  & 6.35  & 9.35  & 0.33 & 0.27 & 0.18  \\
$CC_{t}^{D}$  & 4.98  & 1.05 & 0.36  & 4.50  & 0.63  & 3.35  & 4.95  & 6.71  & 9.73  & 0.68 & 0.52 & 0.36  \\
$RV_{t}^{D-}$ & 4.33  & 1.07 & 0.35  & 4.04  & 0.75  & 2.65  & 4.30  & 6.12  & 9.10  & 0.61 & 0.47 & 0.33  \\
$RV_{t}^{D+}$ & 5.42  & 0.80 & 0.95  & 5.32  & 3.23  & 4.31  & 5.33  & 6.84  & 9.81  & 0.73 & 0.58 & 0.39  \\
$SJ_{t}^{D}$  & 4.49  & 1.82 & -3.08 & 17.48 & -7.10 & 1.38  & 4.75  & 6.32  & 9.40  & 0.11 & 0.10 & 0.07  \\
\midrule
\multicolumn{13}{l}{\textit{Panel B: General attention measures}}                                            \\     
$G_{t}$       & 2.60  & 0.37 & 0.52  & 3.00  & 1.56  & 2.07  & 2.57  & 3.27  & 4.56  & 0.83 & 0.82 & 0.73  \\
$W_{t}$       & 6.26  & 0.87 & -5.37 & 38.11 & 0.00  & 5.88  & 6.31  & 6.91  & 8.30  & 0.65 & 0.49 & 0.30  \\
$T_{t}$       & 6.43  & 0.45 & -0.23 & 4.55  & 5.34  & 5.50  & 6.51  & 7.02  & 9.49  & 0.92 & 0.84 & 0.77  \\
$N_{t}$       & 2.54  & 0.11 & 0.82  & 3.60  & 2.40  & 2.40  & 2.49  & 2.77  & 3.04  & 0.29 & 0.25 & 0.20  \\
\midrule
\multicolumn{13}{l}{\textit{Panel C: General sentiment measures}}                                             \\
$TWP_{t}$     & -2.08 & 0.21 & 0.76  & 5.44  & -3.25 & -2.37 & -2.11 & -1.70 & -0.79 & 0.39 & 0.31 & 0.26  \\
$NPP_{t}$     & -2.25 & 1.16 & 0.02  & 2.12  & -4.99 & -4.08 & -2.27 & -0.24 & -0.06 & 0.35 & 0.06 & 0.02  \\
$TWN_{t}$     & -1.72 & 0.23 & -0.59 & 15.91 & -4.57 & -2.01 & -1.75 & -1.32 & -0.67 & 0.39 & 0.34 & 0.16  \\
$NPN_{t}$     & -1.09 & 1.02 & -1.23 & 3.78  & -4.70 & -3.31 & -0.75 & -0.05 & -0.03 & 0.32 & 0.06 & 0.00       \\
\bottomrule
\end{tabular}

\fnote{Notes: The table reports average, standard deviation, skewness, kurtosis, minimum, 5th, 50th, and 95th percentiles, maximum, and $\rho(.)$, the autocorrelation coefficient of a given order. Panel A includes descriptive statistics of all daily realized measures: realized volatility, jumps, continuous components, semivariances, and signed jumps, in this order. Panel B covers descriptive statistics of general measures of attention (volume of interest) acquired from Google trends, Wikipedia, Twitter, and Newspapers. Panel C reports descriptive statistics of four general sentiment measures: positive sentiment from Twitter, positive sentiment from newspapers, negative sentiment from Twitter, and finally, negative sentiment in news articles. Note that all reported values in Panel A are average values calculated across 404 stocks. Variables included in Pales B and C were aggregated over multiple keywords related to general stock market trading (listed in footnote \ref{fn:kwrds}). In addition, all included variables were log-transformed.}

\end{table}

\subsubsection{Characteristics of the realized measures}

The (log-transformed) realized measures in Panel A of Table \ref{Tbl1} correspond to averages across the $404$ stocks. We observed greater persistence and right-skewness of the overall volatility ($RV_t$). Note that even after $22$ trading days, the average autocorrelation coefficient of volatility is $0.38$, which is much greater than one would expect under an exponential decline ($\approx 0.0007$).


As expected, the jump component ($JC_t^D$) shows lower persistence than the continuous one\footnote{Note that these components correspond to intraday data, whereas the predicted volatility also includes overnight price variation.} ($CC_t^D$), yet the persistence of the jump component is greater than one might expect. However, we record daily jump variation only if it is statistically significant; if the hypothesis of a no-jump day is not rejected, the jump component is equal to $0$. If multiple no-jump days cluster together, we might also observe greater persistence of the jumps.
The signed jump ($SJ_t^D$) behaves differently than the jump component, showing a lower level of persistence. A comparison of the log-transformed semivariances reveals that while both volatilities are highly persistent, the volatility associated with bullish market conditions tends to be less persistent. These results are standard in the literature, but the properties of attention, and particularly sentiment, are less well known.

\subsubsection{Characteristics of attention and sentiment}

From Panels B and C of Table \ref{Tbl1} and from Tables \ref{Tbl1_macro_att} and \ref{Tbl1_macro_sent}, we learn that interest and sentiment toward stock markets and most macroeconomic variables change slowly. The specific results depend on the data source and type of macroeconomic variable.

With respect to attention and sentiment toward the stock market, several relevant findings are notable. \textit{First}, attention from Google Trends ($G_{t}$) and Twitter ($T_{t}$) feeds has similar memory properties, both having autocorrelation coefficients of approximately $0.7$ after $22$ trading days. Attention seems more persistent than volatility. Wikipedia page views ($W_{t}$) show less persistence, and attention derived from news articles ($N_{t}$) appears to behave even more distinctively. \textit{Second}, positive and negative sentiments behave differently as well. Positive tweets seem to be more 'sticky' than negative tweets, which is also true for news articles.

In Tables \ref{Tbl1_macro_att} and \ref{Tbl1_macro_sent}, we breakdown the attention and sentiment measures across the ten macroeconomic news announcement-related topics. We observe considerable variation between Google Trends, Twitter, and newspaper attention measures (Panels A, C, and D of Table \ref{Tbl1_macro_att}), with the most important macroeconomic news announcement-related topics (e.g., GDP, CPI, FOMC meetings) having greater persistence. On the other hand, Wikipedia shows almost identical behavior across news announcement topics. In Table \ref{Tbl1_macro_sent}, we observe that contrary to sentiment related to the stock market (see Table \ref{Tbl1}), positive and negative sentiments associated with the same topics (from the ten macroeconomic news announcements) show similar behavior. For example, positive and negative sentiments related to nonfarm payrolls in news-based articles show similar persistence ($0.87$ and $0.88$), and the same applies to sentiment in Twitter feeds.

\begin{table}[!h]

\centering
\setlength{\tabcolsep}{2.5pt} 
\small

\caption{In-Sample models}
\label{Tbl3_pvals_3}
\begin{tabular}{lrrrrrrrrrrrr}
\toprule
\multicolumn{1}{c}{}  & \multicolumn{1}{c}{HAR} & \multicolumn{1}{c}{** (\%)}  & \multicolumn{1}{c}{CPS} & \multicolumn{1}{c}{HAR-A} & \multicolumn{1}{c}{** (\%)}  & \multicolumn{1}{c}{CPS} & \multicolumn{1}{c}{HAR-S} & \multicolumn{1}{c}{** (\%)}  & \multicolumn{1}{c}{CPS} & \multicolumn{1}{c}{HAR-M} & \multicolumn{1}{c}{** (\%)}  & \multicolumn{1}{c}{CPS} \\
\midrule 
Intercept             & 0.636 & 99.75      & 1.00 & -1.057 & 79.70      & 0.01 & 0.808  & 97.52      & 1.00 & 0.631  & 99.75      & 1.00 \\                   
\multicolumn{13}{l}{\textit{Panel A: Volatility components}}                                   \\
$RV_{t}^{D}$          & 0.474 & 100.00     & 1.00 & 0.449  & 100.00     & 1.00 & 0.457  & 100.00     & 1.00 & 0.470  & 100.00     & 1.00 \\
$RV_{t}^{W}$          & 0.210 & 100.00     & 1.00 & 0.196  & 100.00     & 1.00 & 0.205  & 100.00     & 1.00 & 0.212  & 100.00     & 1.00 \\
$RV_{t}^{M}$          & 0.201 & 100.00     & 1.00 & 0.192  & 100.00     & 1.00 & 0.206  & 100.00     & 1.00 & 0.203  & 100.00     & 1.00 \\
\midrule 
\multicolumn{13}{l}{\textit{Panel B: Attention measures}}                                      \\
$G_{t}$               &       &            &      & 0.190  & 98.27      & 1.00 &        &            &      &        &            &      \\
$W_{t}$               &       &            &      & 0.011  & 10.40      & 0.86 &        &            &      &        &            &      \\
$T_{t}$               &       &            &      & 0.128  & 96.29      & 1.00 &        &            &      &        &            &      \\
$N_{t}$               &       &            &      & 0.239  & 68.32      & 0.98 &        &            &      &        &            &      \\
\midrule 
\multicolumn{13}{l}{\textit{Panel C: Sentiment measures}}                                      \\
$TWP_{t}$             &       &            &      &        &            &      & -0.116 & 66.34      & 0.05 &        &            &      \\
$NPP_{t}$             &       &            &      &        &            &      & -0.009 & 17.08      & 0.17 &        &            &      \\
$TWN_{t}$             &       &            &      &        &            &      & 0.189  & 96.78      & 1.00 &        &            &      \\
$NPN_{t}$             &       &            &      &        &            &      & 0.009  & 14.85      & 0.76 &        &            &      \\
\midrule 
\multicolumn{13}{l}{\textit{Panel D: Macro dummy measures}}                                    \\
$RV_{i,t} \times B_{NFP,t}$  &       &            &       &        &            &       &        &            &       & -0.005 & 13.37     & 0.29 \\
$RV_{i,t} \times B_{IJC,t}$  &       &            &       &        &            &       &        &            &       & 0.001  & 16.09     & 0.48 \\
$RV_{i,t} \times B_{FOMC,t}$ &       &            &       &        &            &       &        &            &       & 0.034  & 79.21     & 0.99 \\
$RV_{i,t} \times B_{GDP,t}$  &       &            &       &        &            &       &        &            &       & 0.003  & 7.43      & 0.62 \\
$RV_{i,t} \times B_{CPI,t}$  &       &            &       &        &            &       &        &            &       & 0.003  & 3.22      & 0.65 \\
$RV_{i,t} \times B_{ISM,t}$  &       &            &       &        &            &       &        &            &       & 0.017  & 50.25     & 0.97 \\
$RV_{i,t} \times B_{SENT,t}$ &       &            &       &        &            &       &        &            &       & -0.011 & 46.04     & 0.06 \\
$RV_{i,t} \times B_{CBCC,t}$ &       &            &       &        &            &       &        &            &       & 0.013  & 31.19     & 0.92 \\
$RV_{i,t} \times B_{RSA,t}$  &       &            &       &        &            &       &        &            &       & -0.004 & 7.92      & 0.29 \\
$RV_{i,t} \times B_{DGO,t}$  &       &            &       &        &            &       &        &            &       & -0.002 & 7.67      & 0.38 \\
\midrule 
\multicolumn{13}{l}{\textit{Model Fit}}                                                        \\
$R^2$                 & 0.571 &            &      & 0.578  &            &      & 0.575  &            &      & 0.577  &            &      \\
$adj. R^2$            & 0.570 &            &      & 0.576  &            &      & 0.574  &            &      & 0.575  &            &     \\
\bottomrule     
\end{tabular}

\fnote{Notes: The table reports average coefficients from HAR model and its extensions, where the averages are taken across models estimated for each of the $404$ stocks. Model parameters were estimated via OLS and used the quadratic spectral kernel heteroskedasticity and autocorrelation consistent estimator with automatic bandwidth selection, following the procedure of \citet{Newey1994}. 
Columns titled '** (\%)' denote the proportion of cases (out of the 404 stocks considered) where the coefficient reached significance on at least the 5\% level.
The CPS denotes a conditionally positive sign, which is a ratio of coefficients with a positive sign. For example, a value of 1.00 means that the corresponding coefficient was positive for all $404$ stocks.
}
\end{table}


\subsection{Attention and sentiment: In-sample volatility models}

In this section, we first explore the importance of including attention and sentiment related to the stock market in the standard HAR model. Next, we augment the HAR model with macroeconomic news announcement dummies. Table \ref{Tbl3_pvals_3} reports the average coefficients from the HAR model and its extensions. For example, the value of $0.474$ in Table \ref{Tbl3_pvals_3} and Panel A is the average coefficient of lagged volatility, which is significant in 100\% of stocks at the 5\% significance level (denoted by **). The coefficient is further associated with a positive sign (CPS) of 1.00, meaning that across all $404$ stocks, the coefficient was positive.

A comparison of the models reveals that attention measures are relevant for most stocks. The log--log specification allows us to compare the magnitude of the coefficient directly. We observe that a $1\%$ change in Google trends leads to a $0.190\%$ increase in volatility, and an equal change in Twitter attention increases volatility by $0.128\%$. These coefficients are positive in $100\%$
of stocks and significant at the $5\%$ level across over $98\%$ and $96\%$ of stocks considered, respectively. 
A larger effect is found for News articles, at $0.239\%$, and a much smaller and rarely significant effect is found for Wikipedia page views, at $0.011\%$. 

In the next model, we study the roles of positive and negative sentiment. Including both improves the fit, and 
negative sentiment tends to increase the volatility level across most of the stocks in the sample, whereas positive sentiment tends to decrease it. The results show that Twitter is more important than newspapers are. The effect for Twitter shows that a $1\%$ increase in sentiment leads to a $-0.116\%$ decrease in volatility for positive tweets and a $0.189\%$ increase in volatility for negative tweets. The corresponding effects for newspaper articles are much smaller, $-0.009\%$ for positive sentiment and $0.009\%$ for negative sentiment. However, the aggregated newspaper indices are rarely significant in an in-sample setting---only in $17.08\%$ of cases for positive sentiment ($NPP_t$) and $14.85\%$ for negative sentiment ($NPN_t$) across all $404$ stocks.


Finally, we augment the HAR with scheduled news announcement dummies (see Equation \ref{eq:har-m}) and obtain the second highest fit overall, suggesting that it might be beneficial to know that there will be a relevant macroeconomic news announcement the next day. FOMC meetings have the greatest effect, which aligns with the literature. However, this is an in-sample analysis, and the model might be overfitted. There might also be a confounding day-of-the-week effect, as certain news announcements tend to be announced on specific days during the week.

\subsection{Attention and sentiment: Out-of-sample volatility models}

This section presents our key results on the basis of an out-of-sample study. We compare the forecasting accuracy of benchmark models with those that include either attention or sentiment related to ten macroeconomic variables.

\subsubsection{The role of attention and sentiment}

Our main results can be found in Table \ref{Tbl4_5w_mse_H1_all}, which shows the out-of-sample performances of the individual models. The reported results are aggregated across 404 stocks. For example, the value $55.45\%$ in the second column of Panel A refers to the proportion of cases (out of $404$ stocks) in which the complete-subset HAR (CSR model) outperforms the benchmark HAR model.

\begin{table}[!ht]

\centering
\setlength{\tabcolsep}{3.5pt} 
\small

\caption{Comparison of forecasting accuracy across $404$ stocks: The mean square error (MSE)}
\label{Tbl4_5w_mse_H1_all}
\small{
\begin{tabular}{l|rrr|rrrr|rrrr}
\toprule
                          & \multicolumn{3}{c}{Benchmark} & \multicolumn{4}{|c}{Attention} & \multicolumn{4}{|c}{Sentiment}     \\
Model                 & \multicolumn{1}{c}{HAR} & \multicolumn{1}{c}{CSR} & \multicolumn{1}{c}{HAR-M} & \multicolumn{1}{|c}{HAR-A} & \multicolumn{1}{c}{CSR-A} & \multicolumn{1}{c}{ALA-A} & \multicolumn{1}{c}{RF-A} & \multicolumn{1}{|c}{HAR-S} & \multicolumn{1}{c}{CSR-S} & \multicolumn{1}{c}{ALA-S} & \multicolumn{1}{c}{RF-S} \\

\midrule
\multicolumn{12}{l}{Panel A: Proportion of cases the column model out-performed the benchmark HAR model}                                                                                                                                                                                                                        \\
                    &                         & 55.45                   & 16.34                     & 35.40                     & 90.59                     & 70.79                     & 70.54                    & 98.51                     & 98.76                     & 94.31                     & 49.01                    \\
\midrule
\multicolumn{12}{l}{Panel B: Average rank of models (1: best-performing model, 11: worst-performing model)}                                                                                                                                                                                                                     \\
                    & 7.80                    & 7.58                    & 9.45                      & 8.58                      & 4.42                      & 5.48                      & 5.49                     & 4.17                      & 2.44                      & 3.05                      & 7.55                     \\
\midrule
\multicolumn{12}{l}{Panel C: Forecast improvements of the column model compared to the benchmark HAR model}                                                                                                                                                                                                                 \\
5\%                 &                         & -11.19                  & -31.36                    & -27.69                    & -3.58                     & -14.95                    & -18.10                   & 1.64                      & 2.40                      & -0.05                     & -27.61                   \\
10\%                &                         & -8.37                   & -22.47                    & -17.60                    & 0.16                      & -7.86                     & -10.12                   & 2.88                      & 4.18                      & 1.85                      & -18.97                   \\
25\%                &                         & -3.20                   & -11.21                    & -9.93                     & 3.31                      & -0.94                     & -1.13                    & 4.79                      & 7.34                      & 5.54                      & -8.46                    \\
50\%                &                         & 0.61                    & -5.87                     & -3.98                     & 7.92                      & 5.92                      & 5.10                     & 7.79                      & 11.54                     & 10.46                     & -0.77                    \\
75\%                &                         & 3.50                    & -1.94                     & 2.83                      & 12.41                     & 12.09                     & 11.33                    & 11.30                     & 16.94                     & 17.31                     & 5.49                     \\
90\%                &                         & 7.16                    & 1.83                      & 8.91                      & 17.89                     & 19.57                     & 20.00                    & 16.23                     & 22.90                     & 23.71                     & 14.25                    \\
95\%                &                         & 10.15                   & 3.83                      & 13.19                     & 23.29                     & 25.79                     & 24.59                    & 18.18                     & 26.62                     & 26.78                     & 19.09                    \\
Mean                &                         & -0.14                   & -9.38                     & -5.94                     & 8.12                      & 5.13                      & 4.61                     & 8.62                      & 12.74                     & 11.79                     & -2.15                    \\
\bottomrule
\end{tabular}
}

\fnote{Notes: The table presents the out-of-sample performances of 11 competing models based on the mean square error losses. The first three columns report the results for three benchmark models, the standard HAR model, the superbenchmark CSR HAR model, and the HAR-M, the standard HAR model expanded with interactions of  $10$ macroeconomic dummy variables with daily realized volatility. The next four models all fall under the description "attention models", although the first model HAR-A only includes general attention, while the next three models (ALA-A, CSR-A, and RF-A) also contain measures of individual, macro-event specific attention measures, only differing in the underlying forecasting methodology. The final four columns cover our four sentiment models, again, with the first model, HAR-S, focused only on general sentiment and the remaining three models also on individual sentiment effects. The averages in Panels A and B and percentiles in Panel C are calculated across $404$ stocks.}

\end{table}

\begin{figure}[ht]
\centering
\caption{Distribution of average (over 404 stocks) percent forecast improvements of four top-performing models: CSR-A (top left), HAR-S (top right), CSR-S (bottom left), and ALA-S (bottom right).}
\label{fig:hist}
\includegraphics[width=\textwidth]{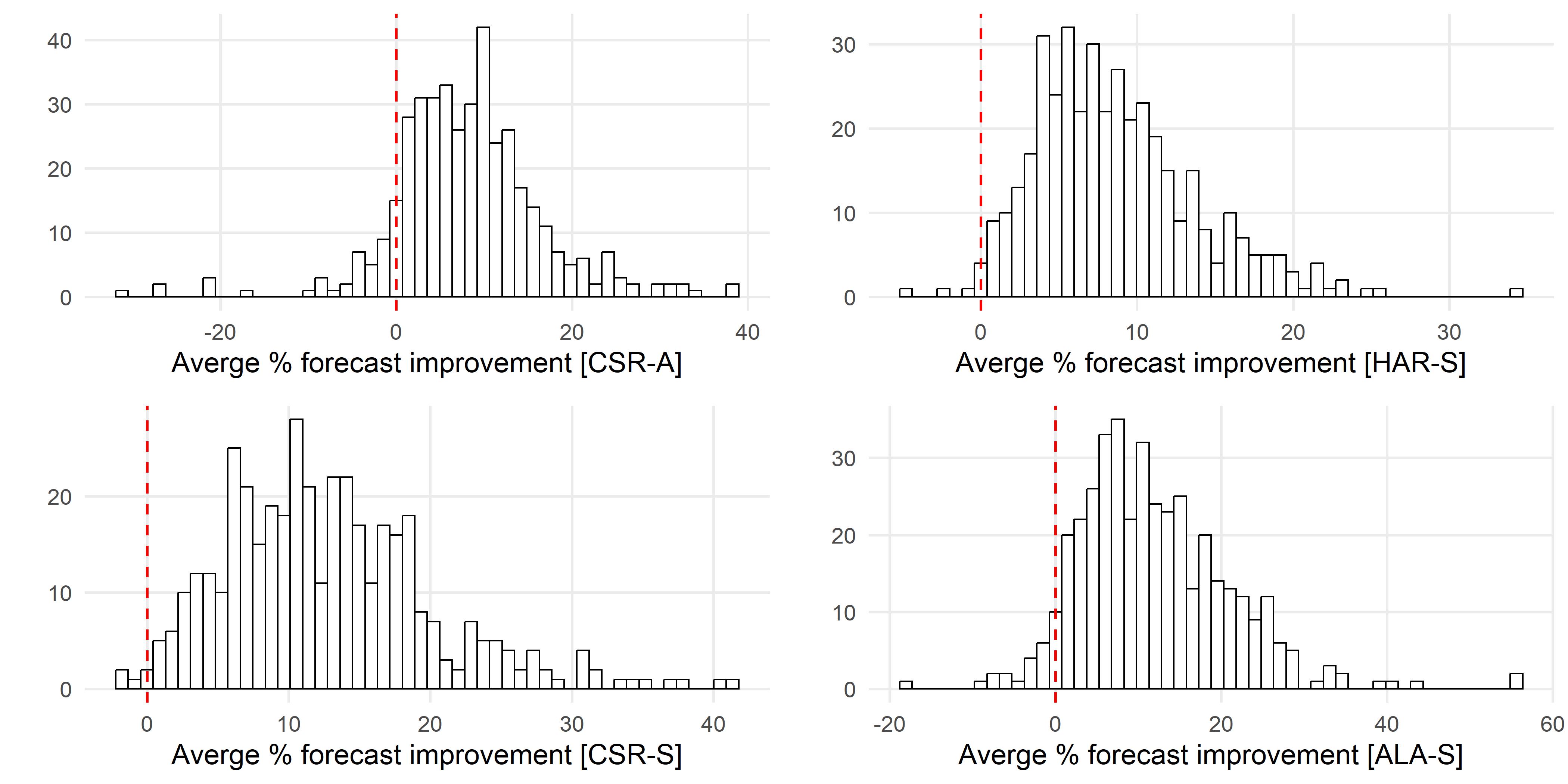}
\end{figure}

In Panel B, the value of $4.42$ in the fifth column indicates the average rank of the column model (CSR-Attention in this case) among all the models. The lower the rank is, the better the model. Note that for 11 models, the lowest possible rank is 1 (meaning that the given model is preferred across all $404$ stocks), and the highest is 11. 
Finally, in Panel C, we report the distribution of potential forecast improvements (in terms of the MSE) with respect to the benchmark HAR model, where positive numbers indicate improved forecast accuracy. For example, the value of $-0.14$ in the last row and second column of Panel C is the average improvement of the complete-subset HAR (CSR HAR) against the HAR model in percentage terms.

Table \ref{Tbl4_5w_mse_H1_all} shows that CSR-Attention outperforms the benchmark HAR model in $90.59\%$ of cases, more than other attention-based models such as adaptive LASSO, RF, and HAR-A, which even underperforms the complete-subset HAR (CSR HAR) model. Among the attention models, the CSR also has the lowest average rank at just $4.42$ (Panel B) and the highest average improvement at $8.12\%$ in terms of MSE against the benchmark HAR model. 
However, we also observe that using attention variables does not necessarily lead to forecast improvements, as the results are model specific. Both RF and adaptive LASSO yield less convincing, although still positive, results.

\begin{table}[!ht]

\centering
\setlength{\tabcolsep}{4pt} 

\caption{Pairwise MCS analysis of the performance of competing models
}
\label{mcs_pairs_0.05_ssm}
\small{
\begin{tabular}{l|rrr|rrrr|rrrr}
\toprule

  & \multicolumn{3}{c}{Benchmark} & \multicolumn{4}{|c}{Attention} & \multicolumn{4}{|c}{Sentiment}     \\
  & \multicolumn{1}{c}{HAR} & \multicolumn{1}{c}{CSR} & \multicolumn{1}{c}{HAR-M} & \multicolumn{1}{|c}{HAR-A} & \multicolumn{1}{c}{CSR-A} & \multicolumn{1}{c}{ALA-A} & \multicolumn{1}{c}{RF-A} & \multicolumn{1}{|c}{HAR-S} & \multicolumn{1}{c}{CSR-S} & \multicolumn{1}{c}{ALA-S} & \multicolumn{1}{c}{RF-S} \\

\midrule
HAR        &          & 0.97    & 0.80     & 0.89   & 1.00  & 0.99  & 0.98 & 1.00   & 1.00  & 1.00  & 0.93 \\
CSR        & 0.93     &         & 0.82     & 0.87   & 1.00  & 0.98  & 1.00 & 1.00   & 1.00  & 1.00  & 0.96 \\
HAR-M      & 1.00     & 1.00    &          & 0.99   & 1.00  & 1.00  & 1.00 & 1.00   & 1.00  & 1.00  & 0.99 \\
\midrule
HAR-A      & 0.95     & 0.98    & 0.93     &        & 1.00  & 0.99  & 0.98 & 1.00   & 1.00  & 1.00  & 0.95 \\
CSR-A      & 0.87     & 0.92    & 0.70     & 0.67   &       & 0.95  & 0.95 & 0.98   & 0.98  & 0.99  & 0.89 \\
ALA-A      & 0.97     & 0.97    & 0.91     & 0.89   & 0.99  &       & 0.99 & 1.00   & 1.00  & 1.00  & 0.95 \\
RF-A       & 0.95     & 0.97    & 0.88     & 0.86   & 0.98  & 0.95  &      & 0.98   & 0.99  & 0.99  & 0.95 \\
\midrule
HAR-S      & 0.74     & 0.84    & 0.56     & 0.82   & 0.97  & 0.95  & 0.93 &        & 0.99  & 0.98  & 0.87 \\
CSR-S      & 0.76     & 0.87    & 0.50     & 0.69   & 0.90  & 0.88  & 0.93 & 0.95   &       & 0.96  & 0.83 \\
ALA-S      & 0.80     & 0.83    & 0.64     & 0.81   & 0.96  & 0.92  & 0.92 & 0.94   & 0.98  &       & 0.82 \\
RF-S       & 0.98     & 0.99    & 0.92     & 0.92   & 0.99  & 0.97  & 0.99 & 0.99   & 1.00  & 1.00  &               \\  
\bottomrule
\end{tabular}
}

\fnote{Notes: The table reports the results of an MCS test of \citet{Hansen2011} (at a $5\%$ confidence level) applied to a pair of competing models: the column model with each of the row models separately. Each reported value stands for the proportion of cases (over $404$ stocks) in which the column model remained in the Superior Set of Models.
}
\vspace{-5pt}
\end{table}


The models that utilize sentiment from tweets and news covering macroeconomic announcements show even more pronounced forecast improvements. Specifically, the sentiment CSR model outperforms HAR in $98.76\%$ of cases, whereas the adaptive LASSO model outperforms HAR in $94.31\%$ of cases. The average ranks are $2.44$ for CSR and $3.05$ for adaptive LASSO. The average improvements are $11.79\%$ for adaptive LASSO and $12.76\%$ for CSR. Surprisingly, the HAR-S model (which includes only sentiment measures related to stock markets) outperforms the HAR model on an even greater percentage of examined stocks, $98.51\%$. 
However, CSR HAR and adaptive LASSO with macroeconomic variable-specific sentiment measures should be preferred, as they are able to provide much greater forecast improvements throughout the whole distribution (across the $404$ stocks), and a similar conclusion can be drawn on the basis of the average rank of the HAR-S model, $4.17$, in contrast to $2.44$ for CSR and $3.05$ for adaptive LASSO.

In Figure \ref{fig:hist}, we present the distribution of forecast improvements across our sample of $404$ stocks for four models, which outperform the benchmark HAR in over $90\%$ of cases: the attention CSR, sentiment HAR, sentiment CSR, and sentiment adaptive LASSO models. The distributions show that for most of the stocks, using attention or sentiment leads to forecast improvements. For many stocks, these improvements are nontrivial.

In an effort to further differentiate the forecasting performance of these models, we study their performance pairwise\footnote{Even though some competing models performed very poorly compared with others (see Panels A, B, and C), it is difficult to statistically distinguish them when all the models are studied together. Specifically, the MCS test on the full set of competing models and across all 404 stocks in the sample, regardless of the level of $\alpha$ considered, suggested that almost all models are in the superior set of models (as termed by \citet{Hansen2011}).} based on (1) the MCS test values, as shown in Table \ref{mcs_pairs_0.05_ssm}, and (2) the one-sided Wilcoxon signed-rank test values, as reported in Table \ref{W_5w_mse_H1_all}. 

Table \ref{mcs_pairs_0.05_ssm} reports the results of a series of pairwise model confidence set (MCS) tests, where each value reported represents a proportion of cases, the column model remained in the superior set of models (SSM). For example, following the first column, the benchmark model HAR holds up very well compared with the other two benchmark models and the general attention model HAR-A (it remains in the SSM in $93\%$, $100\%$, and $95\%$ of cases, respectively) but only survives the MCS test in $70\%$ to $80\%$ of cases with the best-performing models: the complete subset HAR (CSR-S) and the adaptive LASSO HAR (ALA-S), which both employ macroeconomic variable-specific sentiments, as well as the general-sentiment HAR (HAR-S).

\begin{table}[!t]

\centering
\setlength{\tabcolsep}{2.5pt} 
\small

\caption{The average over-performance of models compared to the HAR model based on MSE loss function for individual sectors}
\label{Tbl4_5w_mse_H1}
\hfil 
\begin{tabular}{l|rr|rrrr|rrrr|r}
\toprule
                       & \multicolumn{2}{c|}{Benchmarks} & \multicolumn{4}{c|}{Attention}  & \multicolumn{4}{c|}{Sentiment}    & \multicolumn{1}{l}{} \\
Sectors                & CSR       & HAR-M    & HAR-A    & CSR-A          & ALA-A      & RF-A    & HAR-S       & CSR-S    & ALA-S     & RF-S   & NI \\
\midrule
Industrials            & 64.81          & 18.52    & 22.22    & 96.30          & 81.48      & 83.33   & 100.00      & 100.00   & 94.44     & 62.96  & 54  \\
Health Care            & 68.00          & 22.00    & 14.00    & 86.00          & 58.00      & 82.00   & 100.00      & 98.00    & 96.00     & 56.00  & 50  \\
Information Technology & 67.27          & 10.91    & 9.09     & 83.64          & 69.09      & 87.27   & 98.18       & 98.18    & 98.18     & 70.91  & 55  \\
Communication Services & 64.71          & 5.88     & 17.65    & 94.12          & 41.18      & 64.71   & 100.00      & 100.00   & 88.24     & 47.06  & 17  \\
Consumer Staples       & 53.57          & 14.29    & 35.71    & 75.00          & 53.57      & 67.86   & 100.00      & 100.00   & 92.86     & 32.14  & 28  \\
Consumer Discretionary & 62.75          & 23.53    & 45.10    & 86.27          & 64.71      & 62.75   & 98.04       & 96.08    & 96.08     & 41.18  & 51  \\
Utilities              & 7.41           & 3.70     & 70.37    & 88.89          & 70.37      & 33.33   & 92.59       & 100.00   & 81.48     & 37.04  & 27  \\
Financials             & 45.61          & 12.28    & 49.12    & 98.25          & 85.96      & 64.91   & 98.25       & 100.00   & 94.74     & 29.82  & 57  \\
Materials              & 66.67          & 42.86    & 28.57    & 95.24          & 71.43      & 76.19   & 95.24       & 100.00   & 100.00    & 61.90  & 21  \\
Real Estate            & 33.33          & 7.41     & 85.19    & 100.00         & 77.78      & 51.85   & 100.00      & 96.30    & 92.59     & 33.33  & 27  \\
Energy                 & 52.94          & 17.65    & 41.18    & 100.00         & 94.12      & 76.47   & 100.00      & 100.00   & 94.12     & 58.82  & 17  \\
\midrule
All stocks             & 55.45          & 16.34    & 35.40    & 90.59          & 70.79      & 70.54   & 98.51       & 98.76    & 94.31     & 49.01  & 404 \\
\bottomrule
\end{tabular}

\fnote{Notes: Values in individual columns represent the proportion of cases in which the column model outperformed the benchmark HAR model. These proportions are calculated separately for stocks belonging to $11$ distinct sectors listed in the first column. The last column indicates the number of stocks within each sector.}

\end{table}


Similar results are shown in Table \ref{W_5w_mse_H1_all}. The table contains p values for the one-sided Wilcoxon test after applying the Bonferroni--Holm correction with the alternative hypothesis that the column model outperforms the row model under MSE.
For example, the value in the first row and second column, $0.01$, corresponds to the p value for the test with an alternative hypothesis that the MSE of the complete-subset HAR (CSR HAR) model is significantly lower than the MSE of the benchmark HAR model. Column nine of the Table is of particular interest, as it further proves that the sentiment CSR model outperforms all of the competing models on the basis of the MSE loss function. Closely following is the sentiment model with adaptive LASSO (ALA-S), which outperforms all the considered models except the aforementioned CSR-S model. Behind these two models are the general sentiment model HAR-S and the CSR-A attention model, which is outperformed only by the top three sentiment models. 

Finally, we break down these results across eleven economic sectors. The values in Table \ref{Tbl4_5w_mse_H1} show how many times the given column model outperforms the benchmark HAR model. Event-specific attention and sentiment models consistently improve forecasts across all economic sectors, with the best-performing models showing improvements in over 90\% of the stocks within almost every economic sector (utilities is the only exception for ALA-S). 


\begin{table}[!h]

\centering
\setlength{\tabcolsep}{3.5pt} 
\small

\caption{Comparison of forecasting accuracy across $404$ stocks in top $10\%$ of days with the highest levels in RV: The mean square error (MSE)}
\label{Tbl4_5w_mse_H1_top}
\small{
\begin{tabular}{l|rrr|rrrr|rrrr}
\toprule
                          & \multicolumn{3}{c}{Benchmark} & \multicolumn{4}{|c}{Attention} & \multicolumn{4}{|c}{Sentiment}     \\
Model                 & \multicolumn{1}{c}{HAR} & \multicolumn{1}{c}{CSR} & \multicolumn{1}{c}{HAR-M} & \multicolumn{1}{|c}{HAR-A} & \multicolumn{1}{c}{CSR-A} & \multicolumn{1}{c}{ALA-A} & \multicolumn{1}{c}{RF-A} & \multicolumn{1}{|c}{HAR-S} & \multicolumn{1}{c}{CSR-S} & \multicolumn{1}{c}{ALA-S} & \multicolumn{1}{c}{RF-S} \\
\midrule
\multicolumn{12}{l}{Panel A: Proportion of cases the column model out-performed the benchmark HAR model}                                                                                                                                                                                                                           \\
                    &                         & 51.98                   & 19.55                     & 36.63                     & 89.85                     & 75.74                     & 68.32                    & 98.27                     & 99.26                     & 94.55                     & 45.05                    \\
\midrule
\multicolumn{12}{l}{Panel B: Average rank of models (1: best-performing model, 11: worst-performing model)}                                                                                                                                                                                                                        \\
                    & 7.79                    & 7.74                    & 9.31                      & 8.48                      & 4.29                      & 5.18                      & 5.94                     & 4.20                      & 2.32                      & 2.99                      & 7.76                     \\
\midrule
\multicolumn{12}{l}{Panel C: Forecast improvements of the column model compared to the benchmark HAR model}                                                                                                                                                                                                                    \\
5\%                 &                         & -14.26                  & -34.89                    & -31.69                    & -3.36                     & -16.67                    & -19.77                   & 1.85                      & 3.64                      & -0.23                     & -31.96                   \\
10\%                &                         & -10.13                  & -25.65                    & -17.93                    & 0.03                      & -8.67                     & -12.08                   & 3.49                      & 5.57                      & 2.68                      & -22.33                   \\
25\%                &                         & -3.65                   & -11.71                    & -10.21                    & 4.40                      & 0.21                      & -2.59                    & 5.78                      & 9.46                      & 6.73                      & -9.74                    \\
50\%                &                         & 0.20                    & -6.07                     & -4.06                     & 9.52                      & 7.69                      & 4.84                     & 9.33                      & 14.37                     & 12.78                     & -1.42                    \\
75\%                &                         & 3.38                    & -1.25                     & 3.34                      & 14.84                     & 14.77                     & 12.25                    & 13.20                     & 19.59                     & 19.58                     & 6.38                     \\
90\%                &                         & 7.58                    & 2.60                      & 10.35                     & 20.90                     & 22.40                     & 21.57                    & 17.55                     & 25.65                     & 26.41                     & 16.07                    \\
95\%                &                         & 10.91                   & 4.73                      & 14.21                     & 24.90                     & 27.65                     & 27.06                    & 20.56                     & 29.79                     & 29.94                     & 21.28                    \\
Mean                &                         & -0.56                   & -9.93                     & -6.34                     & 9.64                      & 6.56                      & 4.22                     & 9.89                      & 14.99                     & 13.63                     & -2.90                    \\
\bottomrule
\end{tabular}
}

\fnote{Notes: The table presents the out-of-sample performances of 11 competing models based on the mean square error losses. The first three columns report the results for three benchmark models, the standard HAR model, the superbenchmark CSR HAR model, and the HAR-M, the standard HAR model expanded with interactions of  $10$ macroeconomic dummy variables with daily realized volatility. The next four models all fall under the description "attention models", although the first model HAR-A only includes general attention, while the next three models (ALA-A, CSR-A, and RF-A) also contain measures of individual, macro-event specific attention measures, only differing in the underlying forecasting methodology. The final four columns cover our four sentiment models, again, with the first model, HAR-S focused only on general sentiment and the remaining three models also on individual sentiment effects. The averages in Panels A and B and percentiles in Panel C are calculated across $404$ stocks.}

\end{table}

\begin{table}[!h]

\centering
\setlength{\tabcolsep}{3.5pt} 
\small

\caption{Comparison of forecasting accuracy across $404$ stocks in bottom $90\%$ of days with the lowest levels in realized volatility: The mean square error (MSE)}
\label{Tbl4_5w_mse_H1_bottom}
\small{
\begin{tabular}{l|rrr|rrrr|rrrr}
\toprule
                          & \multicolumn{3}{c}{Benchmark} & \multicolumn{4}{|c}{Attention} & \multicolumn{4}{|c}{Sentiment}     \\
Model                 & \multicolumn{1}{c}{HAR} & \multicolumn{1}{c}{CSR} & \multicolumn{1}{c}{HAR-M} & \multicolumn{1}{|c}{HAR-A} & \multicolumn{1}{c}{CSR-A} & \multicolumn{1}{c}{ALA-A} & \multicolumn{1}{c}{RF-A} & \multicolumn{1}{|c}{HAR-S} & \multicolumn{1}{c}{CSR-S} & \multicolumn{1}{c}{ALA-S} & \multicolumn{1}{c}{RF-S} \\
\midrule
\multicolumn{12}{l}{Panel A: Proportion of cases the column model out-performed the benchmark HAR model}                                                                                                                                                                                                                           \\
                    &                         & 59.41                   & 11.39                     & 38.61                     & 40.59                     & 29.95                     & 69.55                    & 58.91                     & 27.97                     & 38.86                     & 55.20                    \\
\midrule
\multicolumn{12}{l}{Panel B: Average rank of models (1: best-performing model, 11: worst-performing model)}                                                                                                                                                                                                                        \\
                    & 5.30                    & 4.78                    & 8.27                      & 6.75                      & 6.26                      & 7.84                      & 3.70                     & 4.76                      & 6.76                      & 6.29                      & 5.28                     \\
\midrule
\multicolumn{12}{l}{Panel C: Forecast improvements of the column model compared to the benchmark HAR model}                                                                                                                                                                                                                    \\
5\%                 &                         & -5.72                   & -12.66                    & -22.84                    & -12.80                    & -30.28                    & -14.02                   & -3.77                     & -6.18                     & -13.56                    & -21.54                   \\
10\%                &                         & -3.86                   & -10.07                    & -13.82                    & -8.08                     & -20.66                    & -8.12                    & -3.11                     & -5.30                     & -10.08                    & -12.40                   \\
25\%                &                         & -1.39                   & -6.57                     & -7.99                     & -3.60                     & -11.15                    & -2.00                    & -1.20                     & -3.50                     & -5.82                     & -5.31                    \\
50\%                &                         & 0.86                    & -3.98                     & -1.91                     & -0.87                     & -4.62                     & 5.93                     & 0.64                      & -1.54                     & -1.41                     & 1.01                     \\
75\%                &                         & 3.58                    & -2.05                     & 2.06                      & 1.76                      & 0.96                      & 11.29                    & 2.78                      & 0.38                      & 2.46                      & 6.97                     \\
90\%                &                         & 6.85                    & 0.22                      & 6.20                      & 4.13                      & 4.35                      & 17.36                    & 4.87                      & 2.39                      & 6.03                      & 12.87                    \\
95\%                &                         & 8.95                    & 1.07                      & 8.65                      & 5.58                      & 7.86                      & 21.49                    & 6.17                      & 4.41                      & 8.11                      & 16.02                    \\
Mean                &                         & 1.51                    & -5.17                     & -4.58                     & -2.15                     & -7.31                     & 4.62                     & 0.94                      & -1.38                     & -1.40                     & 0.14                     \\
\bottomrule
\end{tabular}
}

\fnote{Notes: The table presents the out-of-sample performances of 11 competing models based on the mean square error losses. The first three columns report the results for three benchmark models, the standard HAR model, the superbenchmark CSR HAR model, and the HAR-M, the standard HAR model expanded with interactions of  $10$ macroeconomic dummy variables with daily realized volatility. The next four models all fall under the description "attention models", although the first model HAR-A only includes general attention, while the next three models (ALA-A, CSR-A, and RF-A) also contain measures of individual, macro-event specific attention measures, only differing in the underlying forecasting methodology. The final four columns cover our four sentiment models, again, with the first model, HAR-S focused only on general sentiment and the remaining three models also on individual sentiment effects. The averages in Panels A and B and percentiles in Panel C are calculated across $404$ stocks. Furthermore, the averages are calculated only on a portion of the test sample, excluding days with the top $10\%$ levels of realized volatility.}

\end{table}


\subsubsection{The role of extreme price variation}
\label{sec:extremes}

Improved volatility forecasts are most in demand during the most volatile market periods. We, therefore, split our main results on the basis of the size of the observed (true) predicted variance. For each stock, given the sample of days for which we had predictions, we select days with the $10\%$ highest overall price variation. Next, we recalculate the forecasting accuracy for all stocks separately for high- and low-volatility days. The results for the high-volatility period are in Table \ref{Tbl4_5w_mse_H1_top}, and the results for low-volatility days are in Table \ref{Tbl4_5w_mse_H1_bottom}.


The results show striking differences between the two volatility periods. During days with high price variation (Table \ref{Tbl4_5w_mse_H1_top}), the attention and sentiment models perform better than when they are evaluated over the full out-of-sample period (Table \ref{Tbl4_5w_mse_H1_all}). 
Improvements are observed mainly in the high outperformance 
of the CSR and adaptive LASSO models (CSR-A, ALA-A, CSR-S, ALA-S), in the average rankings of these models, and in the average forecast improvements. 
For example, the average improvement of the attention CSR model increases from $8.12\%$ to $9.64\%$, whereas those of the sentiment CSR and adaptive LASSO increased from $12.74\%$ and $11.79\%$ to $14.99\%$ and $13.63\%$. 
However, during most trading days, the attention and sentiment models tend to slightly underperform the benchmark HAR model. The sentiment CSR model, on average, reduces forecast accuracy by $-1.38\%$, and the attention CSR model reduces it by even more, $-2.15\%$, whereas among data-intensive models, the attention-based random forest model now performs best but still underperforms the HAR model by $-0.99\%$. 
These results show that using our attention- and sentiment-based models comes at a price, as attention and sentiment work best during periods of highest volatility.  


\subsubsection{Which macroeconomic variables are worth following?}

In this section, we elucidate the role of specific macroeconomic variables. Our analysis is based on two models: i) attention CSR and ii) sentiment CSR. The second model uses both positive and negative variables, which we report separately. In this way, we can also observe and report interactions between macroeconomic news announcements and estimated sentiments.

\begin{table}[ht]

\centering
\setlength{\tabcolsep}{2.5pt} 
\small

\caption{Variable importance analysis from Complete Subset Regression models based on macroeconomic news announcements.}
\label{Tbl6_events_type}

\begin{tabular}{l|rrrrr|rr|rr}
\toprule
Type of variable                 & \multicolumn{5}{c}{Attention}     & \multicolumn{2}{|c}{Positive sentiment}     & \multicolumn{2}{|c}{Negative sentiment}    \\
Source                           & \multicolumn{1}{c}{Bloomberg} & \multicolumn{1}{c}{GT} & \multicolumn{1}{c}{Wikipedia} & \multicolumn{1}{c}{Twitter} & \multicolumn{1}{c}{Newsp.} & \multicolumn{1}{|c}{Twitter} & \multicolumn{1}{c}{Newsp.} & \multicolumn{1}{|c}{Twitter} & \multicolumn{1}{c}{Newsp.} \\
\midrule
NFP     & 3.17     & 2.53     & 2.80     & 3.03     & 3.14     & 3.37     & 3.22     & 4.56     & 5.91     \\
IJC     & 3.51     & 2.86     & 3.00     & 6.26     & 3.13     & 3.33     & 3.74     & 3.36     & 4.69     \\
FOMC    & 5.25     & 4.76     & 3.27     & 6.26     & 4.24     & 3.80     & 3.71     & 3.63     & 3.19     \\
GDP     & 3.08     & 3.06     & 2.63     & 2.73     & 3.05     & 3.38     & 3.87     & 3.86     & 3.57     \\
CPI     & 2.86     & 2.73     & 2.64     & 2.55     & 2.82     & 3.73     & 3.32     & 3.41     & 6.41     \\
ISM     & 4.95     & 2.46     & 2.40     & 2.64     & 2.77     & 3.26     & 5.66     & 4.53     & 3.78     \\
SENT    & 3.40     & 2.45     & 2.29     & 2.43     & 2.73     & 3.21     & 3.29     & 4.12     & 3.60     \\
CBCC    & 4.70     & 2.91     & 2.41     & 2.98     & 2.72     & 4.06     & 3.39     & 2.69     & 2.75     \\
RSA     & 3.16     & 2.52     & 2.38     & 2.96     & 2.80     & 3.41     & 4.08     & 3.79     & 3.63     \\
DGO     & 3.19     & 2.82     & 2.81     & 3.01     & 3.28     & 4.04     & 3.19     & 3.09     & 2.92     \\
\midrule
Stock market     &     & 13.07     & 3.35     & 10.46     & 3.58     & 4.14     & 21.28     & 3.90     & 4.37     \\
\bottomrule
\end{tabular}


\end{table}

Note that in the CSR models, we identify a group of linear models for each trading day (using the discounted mean square error and 5-means clustering; see Section \ref{subsubsec:csr}) that performed the best over the past $500$ out-of-sample volatility forecasts. These models differ with respect to the set of variables they use; they might change daily, and the number of models might change as well. To estimate the importance of a given attention or sentiment variable, we find the percentage of cases in which the given sentiment indicator was used within this group of preferred models\footnote{For example, say that at time $t$, the set of preferred models contained two models, one using attention for FOMC and the other using negative sentiment for GDP-related variables. The next day, $t+1$, the group of preferred models contained one model that used attention in FOMC. The importance of attention to FOMC would have a value of 2/3, and the importance of negative sentiment to GDP would be equal to 1/3.}. The averages are computed across the $404$ stocks as well. For example, the value of $3.17$ in the first row of Table \ref{Tbl6_events_type} means that attention to the announcement of "Change in Nonfarm Payrolls" from Bloomberg was part of the best-performing models in $3.17\%$ of cases; i.e., the higher the number is, the more important the given variable is for forecasting purposes.

The results in Table \ref{Tbl6_events_type} reveal that not all macroeconomic variables are equally relevant. Across multiple data sources, attention to FOMC meetings obtained the highest importance. Other notable variables are the purchasing manager index (ISM), consumer confidence reports (CBCC), and initial jobless claims (IJC). Sentiments show similar results, and high interest is also given to nonfarm payroll reports. Moreover, it appears that for predicting the next day's price variation, sentiments from tweets and newspapers are both relevant sources of data across macroeconomic variables.

In Table \ref{Tbl6_events_type}, we report the importance of the broad stock market-related attention and sentiment variables. As noted in the in-sample analysis, the overall attention and sentiment of the general public toward stock markets are important drivers. The values in the last row are almost always highest in the given column, meaning that among all categories of variables, this broad attention was most often included in the set of best-performing model specifications under the CSR framework. Finally, Table \ref{Tbl6_events_type} also reveals the most useful data sources. For attention, Wikipedia page views have the lowest relevance, whereas Google trends and Twitter have the highest relevance. In terms of sentiment, positive news articles are highly relevant for predicting the next day's price variation. 






\section{Robustness checks}
\label{sec:robust}

\subsection{Alternative construction of volatility estimators}

Our main results are based on the price variation data, which are constructed as a weighted sum of overnight and intraday variation (following \citet{Hansen2005}). To study how this decision influences our results, we performed the same analysis following the simpler approach of \citet{blair2010forecasting}, which does not involve estimating weights, meaning that the intraday and overnight values are simply summed together (weighted equally). Furthermore, our main results employ the RV estimator at a 5-minute sampling frequency as a proxy for intraday variation (see subsection \ref{sec:estimators}). Despite its widespread popularity in the literature and its ability to reduce market microstructure noise effects, other RV estimator alternatives exist. Therefore, we also re-estimate the results via an RV estimator with a 1-minute sampling frequency. 

Overall, the results remain globally unchanged\footnote{To conserve space, the results obtained from all alternative specifications and other robustness checks are available upon request.} under both alternatives---with the nonweighted approach as well as with the 1-minute RV estimator. For instance, the three sentiment specifications (HAR-S, CSR-S, and ALA-S) all retained forecasting improvements in over 90\% of the examined stocks, and the results of the best-performing attention specification (CSR-A) were only slightly lower, with forecasting improvements in approximately 80\% of the stocks, compared with the main results of 90.59\%.

\subsection{Different sentiment analysis techniques}

One of our key results is that sentiment toward key macroeconomic variables has the potential to systematically increase volatility forecasting accuracy for almost all (over 90\%) major U.S. equities. One might ask how strongly our results are driven by the specific choice of the sentiment extraction method \citep{ballinari2021gauge}. To address this issue, we explore two different sentiment analysis techniques, as follows:

\begin{enumerate}
    \item The EmoLex sentiment lexicon of \citep{Mohammad2013}, which contains 14,182 English words mapped into one or more of eight basic emotions (anger, fear, anticipation, trust, surprise, sadness, joy, and disgust). In this way, we classify each tweet or news article according to the extent to which it manifests one of the eight emotions. The emotions that are deemed positive and negative are then aggregated into two corresponding groups. Hence, our set of sentiment indicators contains two types of measures calculated for each search keyword:
    \begin{itemize}
        \item Daily average of positive emotions (combining joy and trust).
        \item Daily average of negative emotions (combining anger, disgust and fear).
    \end{itemize} 
    \item The Valence Aware Dictionary and sEntiment Reasoner (VADER) of \citet{hutto2014vader}, a rule-based lexicon specifically attuned to sentiment in social media contexts. Unlike the EmoLex lexicon, words are directly mapped to positive and negative sentiment values.
\end{enumerate}

Both of these lexicons required a series of additional text-cleaning procedures. These included the translation of emoticons into short word descriptions (using \textcolor{blue}{\hyperlink{http://unicode.org/emoji/charts/full-emoji-list.html}{Unicode.org}}) and the removal of special symbols, website URLs, numbers, punctuation marks, abbreviations, and uppercase letters. Afterward, we ensured that no commonly used words would bias the sentiment measures. Hence, we removed 'SMART' stop words taken from the online appendix of \citep{LewisDavid2004RCV1AN} and included the macroeconomic-variable search keywords. Finally, the text was tokenized---split into smaller units, in our case, words---and indexed with a weighting scheme. We used the well-known bag-of-words (BOW) method, which represents text as a vector of terms weighted according to the frequency of their occurrence in a document.

Finally, these measures were applied in the same way as those estimated via the FinBERT model. The results indicate that the inclusion of event-based sentiment measures leads to forecasting improvements regardless of the sentiment extraction method applied. The CSR-S model consistently improves volatility forecasts for over $90\%$ of stocks across all sentiment techniques ($94.55\%$ with VADER, $95.30\%$ with EmoLex and $98.76\%$ with FinBERT). The lexicon-based methods lead to slightly weaker results with the adaptive LASSO model---there are forecasting improvements in $87.87\%$ of stocks with ALA-S VADER and $89.85\%$ with ALA-S EmoLex, compared with $94.31\%$ with the FinBERT model. 

We also found substantially worse performance by the general sentiment model HAR-S - $30.69\%$ with the VADER lexicon and $77.97\%$ with the EmoLex lexicon. This section thus reveals that the choice of the sentiment extraction method matters, and to achieve more consistent forecast improvement (across U.S. equities), one needs either (i) more advanced sentiment analysis techniques, such as the FinBERT model, or (ii) the addition of macroeconomic variable-specific measures.

\subsection{Multiple-day-ahead forecasting}

Our main results show that the attention and sentiment measures related to both the general stock market and macroeconomic variables can significantly improve the predictions of the benchmark HAR model in one-day-ahead volatility predictions. As part of this section, we also explore the prediction power of these measures over longer forecasting horizons, particularly for 5- and 22-day-ahead predictions, which are the usual choices in the literature. The forecasting accuracies of all eleven competing models decrease with increasing forecasting horizon, which is a typical result in the literature. Interestingly, both the CSR attention and the CSR sentiment models still seem to improve the forecasting accuracy for both the 5-day- and 22-day-ahead predictions. The third model, which seems to consistently outperform the benchmark HAR model across forecasting horizons, is the complete-subset HAR (CSR HAR) model. This finding further supports our conclusions in Section \ref{sec:extremes} that a combination of a sentiment or attention volatility forecast and the complete-subset HAR model has the potential to yield the best results. 



\subsection{Results under alternative loss functions}
\label{subsubsec:losses}

Finally, we explore out-of-sample forecast accuracy under three other well-known loss functions: the empirical quasilikelihood (QLIKE), which penalizes volatility underestimation more strictly; the mean absolute error (MAE); and the mean absolute percentage error (MAPE). These are defined in Equations \ref{eq:qlike}, \ref{eq:mae} and \ref{eq:mape}, respectively.

\begin{equation}
\label{eq:qlike}
    QLIKE=n^{-1}\sum_{t=1}^{n} \frac{RV_{t+1}}{\widehat{RV}_{t+1}}-\ln\left({\frac{RV_{t+1}}{\widehat{RV}_{t+1}}}\right)-1
\end{equation}

\begin{equation}
\label{eq:mae}
    MAE=n^{-1}\sum_{t=1}^{n} \left|RV_{t+1}-\widehat{RV}_{t+1}\right|
\end{equation}

\begin{equation}
\label{eq:mape}
    MAPE=n^{-1}\sum_{t=1}^{n} \left|\frac{RV_{t+1}-\widehat{RV}_{t+1}}{RV_{t+1}}\right|
\end{equation}

The rankings of most models tend to change, yet the CSR-A and CSR-S models remain in the top rankings. These two models are built within the CSR framework and utilize investor sentiment or attention to the stock market as well as macroeconomic variables. Thus, the main conclusions remained unchanged. 

More specifically, the MAE loss function indicates the same four top models as the mean square error. These models outperform the HAR benchmark model on $88.86\%$ of stocks with CSR-A, $95.30\%$ with HAR-S, $98.76$ with CSR-S and $83.42\%$ with the ALA-S model, with average forecasting improvements ranging from $1.59\%$ to $2.56\%$. The MAPE also confirms the predictive abilities of the CSR-S and CSR-A models, with forecasting improvements of $1.03\%$ and $0.39\%$, beating the HAR benchmark on $90.53\%$ and $66.34\%$ of stocks, respectively. The MAPE, however, also favors macroeconomic variable-specific attention and sentiment models built with the random forest (RF-A and RF-S), whose forecasting improvements compared with the HAR benchmark average approximately $5.24\%$ and $5.69\%$, outperforming the benchmark on $97.03\%$ and $96.53\%$ of stocks. Finally, if we consider the forecasting accuracy based on the QLIKE loss criterion, the best-performing model is HAR-A, which only uses attention to the stock market, with average improvements of $1.04\%$, surpassing the benchmark HAR on $80.45\%$ of stocks. Nonetheless, the second and third places are once again taken by the two complete-subset regression models---CSR-A and CSR-S---outperforming the benchmark model on $66.83\%$ and $62.87\%$ of stocks, respectively. These results show that the choice of model and loss function matter; our main results hold for symmetric loss functions and complete subset models that employ attention and sentiment measures.

\section{Conclusion}

While major macroeconomic variables are known to be of interest to equity market participants, it is not clear how such variables should be used for forecasting purposes; 
in our setup for volatility forecasting.
Several studies have focused on the scheduled announcements of key macroeconomic variables; however, such an approach is not suitable for forecasting purposes, as i) announcements are not frequent, ii) the exact timing of the announcement might change, iii) the market expectations about the announcement vary over time, iv) the news might be priced in over prolonged periods, and v) the importance of a given news announcement might change over time. In this study, we argue that a suitable proxy for the role of macroeconomic variables is the attention and sentiment related to ten of these macroeconomic variables. 

We focus on ten major macroeconomic variables and day-ahead volatility predictions of individual U.S. equities. We set up four models with attention measures and four models with sentiment and test them against the results of three benchmark models. These include the standard HAR model of \citet{corsi2009simple}, a complete-subset HAR model (CSR HAR) based on a complete-subset regression combination of other well-known HAR extensions, and a HAR model augmented with macroeconomic dummy variables. We further distinguish between attention \& sentiment toward general stock market trading and measures specific to macroeconomic news announcements. We employ three different modeling techniques: adaptive LASSO \citep{zou2006adaptive}, random forest by \citet{breiman2001random}, and complete subset regression by \citet{elliott2013complete}.

Overall, the results show that sentiment related to macroeconomic news announcements systematically leads to improved volatility forecasts, as evaluated via the MSE criterion. Within the complete-subset HAR framework, attention and sentiment prove helpful for $90.59\%$ and $98.76\%$ of stocks, respectively. Similar results are achieved with adaptive LASSO ($94.31\%$) using the sentiment measures as well as with the HAR-S model---the benchmark HAR model augmented with sentiment toward the stock market in general ($98.51\%$ of stocks, although with lower forecast improvements on average). For the best models, the forecasting improvements range from $8.12\%$ to $12.74\%$ on average and are found in every economic sector. Interestingly, after splitting these results into predictions for high-volatility and low-volatility days, we discovered that the most successful sentiment \& attention models 
are particularly useful for days of extreme price variation. During days with lower volatility, other models (such as CSR HAR) appear to produce more consistent and reliable forecasts. This suggests that further developments combining the benchmark CSR HAR model during calm periods with models that utilize sentiment and/or attention measures during periods of turmoil might be the ideal combination; this is left for further research. 

We find that the most relevant macroeconomic news announcements are FOMC meetings (consistent with the literature) and labor market indicators. The most important data sources are Google Trends, Twitter, and newspaper articles, mainly general attention towards the stock market from Google Trends and Twitter, as well as event-specific sentiment from both Twitter and newspapers and general positive sentiment toward the stock market from news articles.

Finally, we ran a battery of robustness tests and verified that the main conclusions of this study hold under most scenarios. These include (i) alternative specifications of the RV estimator, such as applying it at a 1-minute sampling frequency and taking the simpler approach to combining overnight and intraday variation (following \citet{blair2010forecasting}), (ii) different sentiment extraction techniques (the EmoLex lexicon of \citet{Mohammad2013} and VADER lexicon of \citet{hutto2014vader}), (iii) multiple-day-ahead forecast horizons, namely, 5- and 22-day-ahead horizons, and (iv) forecast evaluation under three other loss functions---the QLIKE, MAE, and MAPE---although for the QLIKE, the results were less impressive.

\section*{Data availability}
Data will be made available on request.

 \bibliographystyle{elsarticle-harv} 
 
{\small
\setstretch{1.5}
\setlength{\bibsep}{1.2pt}
\bibliography{cas-refs}
}

\clearpage
\appendix

\section{Appendix}

\label{sec:appendix}

\setcounter{table}{0}
\setcounter{figure}{0} 


\begin{table}[ht]

\centering
\setlength{\tabcolsep}{2.5pt} 
\renewcommand{\arraystretch}{1.1} 
\footnotesize

\caption{Macroeconomic news announcements keywords}
\label{Tbl_event_small}

\begin{tabular}{ll}
\toprule
Macroeconomic Event                           & Keywords                                                                                                                                                                                                                                                               \\
\midrule
Change in Nonfarm Payrolls      & \multicolumn{1}{m{12cm}}{Nonfarm Payrolls, Nonfarm Payroll, Nonfarm   Payroll Report, Non Farm Payroll Data, Non Farm Payroll Dates, NFP Report,   Nonfarm Employment, Current Employment statistics, Employment Situation,   Employment Situation Summary, Employment Situation Report}         \\
\midrule
Initial Jobless Claims          & \multicolumn{1}{m{12cm}}{Initial Claims, Initial Claims Data, Jobless   Claims, Jobless Claims Report, Initial Jobless Claims, Weekly Jobless Claims,   Unemployement, Unemployment Claims, Unemployment Benefits Claims}                                                                        \\
\midrule
FOMC Rate Decision              & \multicolumn{1}{m{12cm}}{Federal Funds Rate, Rate Decision, Interest   Rate Decision, Interest Rate Increase, Monetary Policy, FOMC, FOMC Meeting,   FOMC Rate Decision, Fed Funds Rate, Fed Rate Hike, Fed Raise Rates, Interest   Rates Fed}                                                   \\
\midrule
GDP Annualized QoQ              & \multicolumn{1}{m{12cm}}{GDP, GDP Growth, GDP Data, GDP Release, Gross   Domestic Product, Domestic Product, US GDP, GDP Quarter, Change In GDP, GDP   Annual Rate, Yearly GDP, Economic Activity, Economic Growth}                                                                              \\
\midrule
CPI MoM                         & \multicolumn{1}{m{12cm}}{CPI, CPI Rate, CPI Index, CPI Data, CPI   Forecast, Monthly CPI, BLS CPI, Consumer Prices, Consumer Price Index,   Inflation Forecast, Inflation, Inflation Rate}                                                                                                       \\
\midrule
ISM Manufacturing               & \multicolumn{1}{m{12cm}}{ISM Manufacturing, ISM PMI, ISM Report, ISM   Survey, ISM Index, PMI Number, PMI Index, PMI Manufacturing, Manufacturing   Index, Purchasing Managers Index, ISM Report On Business,  Institute For Supply Management}                                                  \\
\midrule
U. of Mich. Sentiment           & \multicolumn{1}{m{12cm}}{Consumer Sentiment, Sentiment Index, Consumer   Sentiment Index, Michigan Consumer Sentiment, Michigan Consumer Sentiment   Index, Consumer Sentiment Data, University Of Michigan Consumer Sentiment   Index}                                                          \\
\midrule
Conf. Board Consumer Confidence & \multicolumn{1}{m{12cm}}{Consumer Confidence, Consumer Confidence   Report, Consumer Confidence Index, Consumer Confidence Data, Consumer   Confidence Survey, Conference Board, Conference Board Index, Conference Board   Consumer Confidence, Conference Board Confidence, Consumer Optimism} \\
\midrule
Retail Sales Advance MoM        & \multicolumn{1}{m{12cm}}{Retail Sales Advance, Sales Advance, Retail   Advance, Retail Trade, Retail Sales Report, Monthly Retail Trade, Advance   Monthly Sales}                                                                                                                                \\
\midrule
Durable Goods Orders            & \multicolumn{1}{m{12cm}}{Durable Goods Orders, Durable Goods Data,   Durable Goods Index, Goods Orders Data, Durable Goods Report, Long-term goods}                                                                                                                                              \\
\bottomrule
\end{tabular}


\end{table}
\begin{table}[ht]

\centering
\setlength{\tabcolsep}{2.5pt} 
\renewcommand{\arraystretch}{1.1} 
\footnotesize

\caption{Wikipedia-specific keywords: titles of the Wikipedia pages}
\label{Tbl_wiki_keywords}

\begin{tabular}{ll}
\toprule
Macroeconomic Event                           & Keywords                                                                                                                                                                                                                                                               \\
\midrule
Change in Nonfarm Payrolls      & \multicolumn{1}{m{12cm}}{Nonfarm Payrolls, Labor market, Economic indicator, Unemployment, Employment Situation Summary}         \\
\midrule
Initial Jobless Claims          & \multicolumn{1}{m{12cm}}{Jobless claims, Initial Jobless Claims, Unemployment benefits, Unemployement}                                                                        \\
\midrule
FOMC Rate Decision              & \multicolumn{1}{m{12cm}}{Federal Funds Rate, Interest rate, Federal Reserve, Federal Open Market Committee, Federal funds, Monetary Policy}                                                   \\
\midrule
GDP Annualized QoQ              & \multicolumn{1}{m{12cm}}{GDP, Gross Domestic Product, Final goods, Economic growth, Domestic Product, US GDP, Economy}                                                                              \\
\midrule
CPI MoM                         & \multicolumn{1}{m{12cm}}{CPI, Consumer Price Index, United States Consumer Price Index, Price index, Inflation, Market basket}                                                                                                       \\
\midrule
ISM Manufacturing               & \multicolumn{1}{m{12cm}}{ISM Report On Business, ISM Manufacturing Index, Institute for Supply Management, Purchasing Managers' Index}                                                  \\
\midrule
U. of Mich. Sentiment           & \multicolumn{1}{m{12cm}}{University Of Michigan Consumer Sentiment Index, Consumer confidence, University of Michigan, Leading Indicator Composite Index, United States Department of Commerce, Bureau of Economic Analysis, Consumer Sentiment, Consumer confidence Index}                                                          \\
\midrule
Conf. Board Consumer Confidence & \multicolumn{1}{m{12cm}}{Consumer confidence Index, Economic indicator, Consumption, Consumer confidence, The Conference Board} \\
\midrule
Retail Sales Advance MoM        & \multicolumn{1}{m{12cm}}{Retail}                                                                                                                                \\
\midrule
Durable Goods Orders            & \multicolumn{1}{m{12cm}}{Monthly Full Report on Manufacturers' Shipments, Inventories, and Orders, Durable good, United States Census Bureau}        \\
\midrule
\midrule
General keywords                        & \multicolumn{1}{m{12cm}}{Stock Market, Stocks, Bullish, Bearish, S\&P 500, Wall Street, financial markets, wall street journal, nasdaq, nyse, earnings per share,  quarterly report, earnings call,  price–earnings ratio,  price to book,  market capitalization, market price, financial times, VIX, market volatility, gold price, t bill, treasury bill, treasury bond, 401(k), Asset Allocation, pension fund, trading volume, bear market, bull market, day trading, technical analysis, dividend yield, futures contract, Google Finance,  Yahoo! Finance,  marketwatch, hedge fund, stock market index, mutual fund, economic recession, stop order, limit order, trading strategy, yield curve, option contract, stock symbol, market order, penny stocks, market bubble, financial crisis, market liquidity, The Motley Fool, Bloomberg.com, seeking alpha, fidelity investments, etrade, ameritrade, Implied volatility, FTSE 100 Index, Nikkei 225, Hang Seng Index, EURO Stoxx 50, Russell 2000, European Central Bank, Eurodollar, Robinhood Markets}        \\
\bottomrule
\end{tabular}


\end{table}

\begin{figure}[h]
\centering
\caption{Macronews-specific attention measures around announcements}
\label{fig:macroatt}
\includegraphics[width=\textwidth]{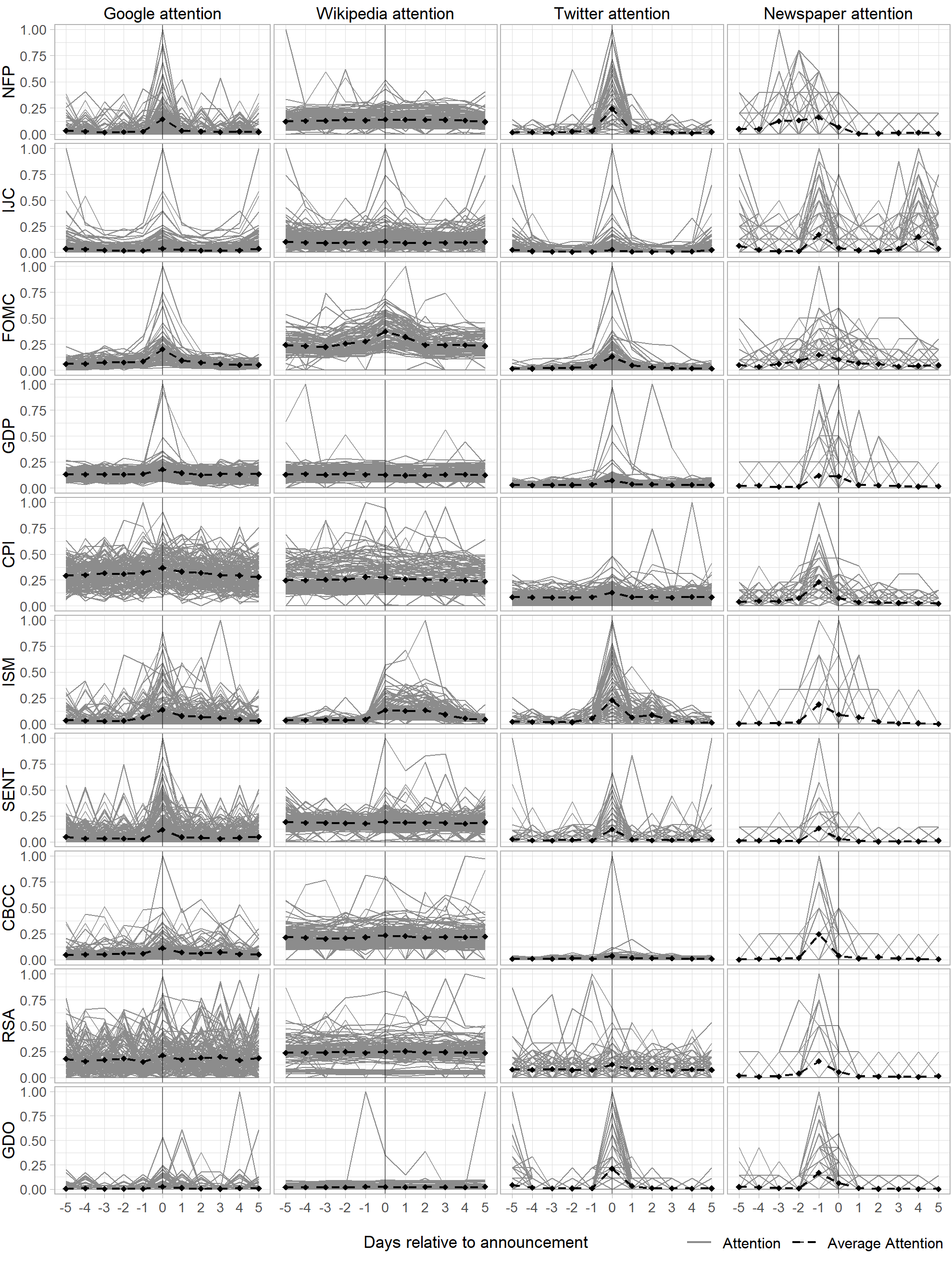}
\end{figure}

\begin{table}[ht]

\centering
\renewcommand{\arraystretch}{1.1} 
\scriptsize

\caption{Descriptive statistics of macro-news specific attention measures.}
\label{Tbl1_macro_att}
\begin{tabular}{lrrrrrrrrrrrr}
\toprule
Variable     & Mean & S.D. & Skew. & Kurt. & Min. & 5th  & 50th & 95th & Max.  & $\rho(1)$ & $\rho(5)$ & $\rho(22)$ \\
\midrule
\multicolumn{13}{l}{\textit{Panel A: Google Trends attention measures}}   \\
$G_{NFP,t}$  & 0.60 & 1.09 & 1.62  & 4.21  & 0.00 & 0.00 & 0.00 & 3.18 & 4.79  & 0.06      & 0.03      & 0.00       \\
$G_{IJC,t}$  & 1.84 & 0.84 & 0.36  & 3.42  & 0.00 & 0.51 & 1.83 & 3.24 & 5.93  & 0.59      & 0.68      & 0.47       \\
$G_{FOMC,t}$ & 2.17 & 0.64 & -0.04 & 3.40  & 0.00 & 1.11 & 2.19 & 3.18 & 5.04  & 0.57      & 0.39      & 0.18       \\
$G_{GDP,t}$  & 2.90 & 0.39 & -0.67 & 4.53  & 0.48 & 2.19 & 2.97 & 3.41 & 4.90  & 0.65      & 0.56      & 0.23       \\
$G_{CPI,t}$  & 3.36 & 0.33 & -0.29 & 2.98  & 2.18 & 2.80 & 3.39 & 3.87 & 4.36  & 0.45      & 0.40      & 0.21       \\
$G_{ISM,t}$  & 0.63 & 0.84 & 1.36  & 3.93  & 0.00 & 0.00 & 0.22 & 2.46 & 3.92  & 0.24      & 0.12      & 0.25       \\
$G_{SENT,t}$ & 0.83 & 1.11 & 1.01  & 2.72  & 0.00 & 0.00 & 0.00 & 3.04 & 4.49  & 0.06      & 0.12      & 0.03       \\
$G_{CBCC,t}$ & 1.31 & 0.70 & 0.34  & 3.04  & 0.00 & 0.25 & 1.32 & 2.51 & 4.20  & 0.33      & 0.32      & 0.29       \\
$G_{RSA,t}$  & 2.10 & 1.16 & -0.72 & 2.48  & 0.00 & 0.00 & 2.40 & 3.58 & 4.36  & 0.03      & 0.06      & 0.01       \\
$G_{DGO,t}$  & 0.14 & 0.56 & 4.19  & 20.47 & 0.00 & 0.00 & 0.00 & 1.59 & 4.60  & -0.01     & -0.01     & 0.02       \\
\midrule
\multicolumn{13}{l}{\textit{Panel B: Wikipedia attention measures}}   \\
$W_{NFP,t}$  & 6.65 & 1.31 & -4.26 & 22.09 & 0.00 & 6.01 & 6.79 & 7.62 & 8.98  & 0.74      & 0.68      & 0.53       \\
$W_{IJC,t}$  & 5.87 & 1.25 & -3.84 & 18.66 & 0.00 & 5.37 & 5.98 & 6.83 & 8.55  & 0.73      & 0.68      & 0.55       \\
$W_{FOMC,t}$ & 6.07 & 1.19 & -4.38 & 22.66 & 0.00 & 5.58 & 6.24 & 6.91 & 8.04  & 0.71      & 0.65      & 0.50       \\
$W_{GDP,t}$  & 6.51 & 1.25 & -4.58 & 24.25 & 0.00 & 6.09 & 6.64 & 7.33 & 8.81  & 0.70      & 0.64      & 0.48       \\
$W_{CPI,t}$  & 6.57 & 1.27 & -4.13 & 21.39 & 0.00 & 6.04 & 6.70 & 7.55 & 8.27  & 0.72      & 0.67      & 0.53       \\
$W_{ISM,t}$  & 3.79 & 0.94 & -1.93 & 10.88 & 0.00 & 2.95 & 3.81 & 5.18 & 7.05  & 0.68      & 0.39      & 0.47       \\
$W_{SENT,t}$ & 6.34 & 1.14 & -4.93 & 27.37 & 0.00 & 5.97 & 6.54 & 6.93 & 8.25  & 0.69      & 0.65      & 0.50       \\
$W_{CBCC,t}$ & 5.60 & 1.05 & -4.46 & 24.36 & 0.00 & 5.09 & 5.76 & 6.37 & 7.42  & 0.64      & 0.61      & 0.39       \\
$W_{RSA,t}$  & 7.35 & 1.58 & -3.87 & 18.10 & 0.00 & 5.83 & 7.78 & 8.29 & 9.21  & 0.75      & 0.71      & 0.57       \\
$W_{DGO,t}$  & 6.63 & 1.31 & -4.05 & 21.48 & 0.00 & 6.07 & 6.73 & 8.03 & 10.81 & 0.75      & 0.68      & 0.53       \\
\midrule
\multicolumn{13}{l}{\textit{Panel C: Twitter attention measures}}   \\
$T_{NFP,t}$  & 0.87 & 0.34 & 2.69  & 12.30 & 0.69 & 0.69 & 0.69 & 1.39 & 3.14  & 0.16      & 0.06      & 0.05       \\
$T_{IJC,t}$  & 1.34 & 0.74 & 1.52  & 5.87  & 0.69 & 0.69 & 1.10 & 2.78 & 6.10  & 0.35      & 0.68      & 0.10       \\
$T_{FOMC,t}$ & 1.49 & 0.95 & 0.25  & 3.01  & 0.00 & 0.00 & 1.61 & 3.05 & 5.61  & 0.68      & 0.54      & 0.49       \\
$T_{GDP,t}$  & 3.47 & 0.61 & 0.12  & 3.71  & 2.20 & 2.40 & 3.53 & 4.36 & 6.84  & 0.81      & 0.73      & 0.69       \\
$T_{CPI,t}$  & 3.30 & 0.59 & -0.38 & 2.88  & 1.95 & 2.20 & 3.40 & 4.14 & 5.70  & 0.81      & 0.76      & 0.71       \\
$T_{ISM,t}$  & 0.44 & 0.62 & 1.50  & 5.18  & 0.00 & 0.00 & 0.00 & 1.61 & 3.40  & 0.34      & 0.06      & 0.33       \\
$T_{SENT,t}$ & 0.25 & 0.46 & 1.89  & 6.56  & 0.00 & 0.00 & 0.00 & 1.10 & 2.94  & 0.05      & 0.05      & 0.01       \\
$T_{CBCC,t}$ & 1.35 & 0.77 & -0.12 & 2.80  & 0.00 & 0.00 & 1.39 & 2.49 & 5.74  & 0.52      & 0.45      & 0.41       \\
$T_{RSA,t}$  & 0.61 & 0.56 & 0.50  & 2.76  & 0.00 & 0.00 & 0.69 & 1.61 & 3.58  & 0.24      & 0.22      & 0.09       \\
$T_{DGO,t}$  & 0.11 & 0.33 & 3.54  & 16.52 & 0.00 & 0.00 & 0.00 & 0.69 & 2.30  & 0.07      & -0.03     & 0.10       \\
\midrule
\multicolumn{13}{l}{\textit{Panel D: News articles attention measures}}   \\
$N_{NFP,t}$  & 1.15 & 0.13 & 2.66  & 10.41 & 1.10 & 1.10 & 1.10 & 1.39 & 2.08  & 0.29      & -0.03     & 0.13       \\
$N_{IJC,t}$  & 0.23 & 0.43 & 1.71  & 4.98  & 0.00 & 0.00 & 0.00 & 1.10 & 2.20  & 0.09      & 0.51      & -0.06      \\
$N_{FOMC,t}$ & 1.49 & 0.17 & 1.83  & 6.55  & 1.39 & 1.39 & 1.39 & 1.79 & 2.64  & 0.37      & 0.20      & 0.14       \\
$N_{GDP,t}$  & 0.08 & 0.24 & 2.96  & 11.12 & 0.00 & 0.00 & 0.00 & 0.69 & 1.61  & 0.13      & 0.04      & 0.02       \\
$N_{CPI,t}$  & 1.50 & 0.18 & 1.82  & 7.27  & 1.39 & 1.39 & 1.39 & 1.79 & 2.83  & 0.27      & 0.03      & 0.14       \\
$N_{ISM,t}$  & 0.05 & 0.18 & 4.02  & 18.55 & 0.00 & 0.00 & 0.00 & 0.69 & 1.39  & 0.20      & -0.05     & 0.12       \\
$N_{SENT,t}$ & 0.09 & 0.26 & 2.97  & 11.47 & 0.00 & 0.00 & 0.00 & 0.69 & 2.08  & 0.07      & -0.07     & -0.04      \\
$N_{CBCC,t}$ & 0.73 & 0.13 & 4.07  & 20.76 & 0.69 & 0.69 & 0.69 & 1.10 & 1.79  & 0.09      & -0.02     & -0.02      \\
$N_{RSA,t}$  & 0.06 & 0.21 & 3.24  & 12.77 & 0.00 & 0.00 & 0.00 & 0.69 & 1.61  & 0.08      & 0.02      & 0.08       \\
$N_{DGO,t}$  & 0.75 & 0.18 & 3.80  & 18.61 & 0.69 & 0.69 & 0.69 & 1.10 & 2.20  & 0.23      & -0.06     & 0.20       \\
\bottomrule
\end{tabular}

\fnote{Notes: The table reports descriptive statistics (average, standard deviation, skewness, kurtosis, minimum, 5th, 50th, and 95th percentiles, maximum, and $\rho(.)$, the autocorrelation coefficient of a given order) of attention measures (volume of interest) related to $10$ specific macroeconomic news announcements. Panels A-D report on daily attention measures acquired from Google trends, Wikipedia, Twitter, and Newspapers, in this order. Note that all included variables were aggregated over multiple keywords related to each macroeconomic news announcement (listed in Table \ref{Tbl_event_small}) and subsequently log-transformed.}

\end{table}

\begin{table}[ht]

\centering
\renewcommand{\arraystretch}{1.1} 
\scriptsize

\caption{Descriptive statistics of macro-news specific sentiment measures.}
\label{Tbl1_macro_sent}
\begin{tabular}{lrrrrrrrrrrrr}
\toprule
Variable          & Mean & S.D. & Skew. & Kurt. & Min. & 5th  & 50th & 95th & Max.  & $\rho(1)$ & $\rho(5)$ & $\rho(22)$ \\
\midrule
\multicolumn{13}{l}{\textit{Panel A: Twitter positive sentiment measures}}   \\
$TWP_{NFP,t}$  & 0.08  & 0.20 & 1.68  & 4.74  & -0.12 & -0.09 & -0.01 & 0.60  & 0.65  & 0.69      & 0.22      & 0.00       \\
$TWP_{IJC,t}$  & 0.04  & 0.15 & 2.17  & 8.15  & -0.13 & -0.10 & -0.01 & 0.34  & 0.64  & 0.45      & 0.08      & 0.01       \\
$TWP_{FOMC,t}$ & 0.09  & 0.13 & 1.51  & 5.92  & -0.14 & -0.07 & 0.07  & 0.34  & 0.65  & 0.27      & 0.09      & 0.02       \\
$TWP_{GDP,t}$  & 0.18  & 0.10 & 1.16  & 6.78  & -0.13 & 0.03  & 0.18  & 0.37  & 0.65  & 0.27      & 0.15      & 0.13       \\
$TWP_{CPI,t}$  & 0.10  & 0.09 & 2.00  & 11.03 & -0.14 & -0.01 & 0.09  & 0.24  & 0.64  & 0.15      & 0.10      & 0.02       \\
$TWP_{ISM,t}$  & 0.14  & 0.25 & 0.98  & 2.52  & -0.12 & -0.10 & 0.02  & 0.64  & 0.65  & 0.64      & 0.24      & 0.08       \\
$TWP_{SENT,t}$ & 0.18  & 0.27 & 0.75  & 1.95  & -0.11 & -0.09 & 0.05  & 0.64  & 0.65  & 0.74      & 0.28      & 0.05       \\
$TWP_{CBCC,t}$ & 0.13  & 0.18 & 1.01  & 3.39  & -0.15 & -0.08 & 0.09  & 0.50  & 0.65  & 0.26      & 0.11      & -0.01      \\
$TWP_{RSA,t}$  & 0.08  & 0.18 & 1.81  & 5.33  & -0.11 & -0.06 & 0.01  & 0.54  & 0.66  & 0.47      & 0.05      & 0.03       \\
$TWP_{DGO,t}$  & 0.19  & 0.28 & 0.62  & 1.77  & -0.12 & -0.10 & 0.06  & 0.65  & 0.65  & 0.90      & 0.60      & 0.03       \\
\midrule
\multicolumn{13}{l}{\textit{Panel B: News articles positive sentiment measures}}   \\
$TWN_{NFP,t}$  & 0.27  & 0.25 & 0.19  & 1.69  & -0.12 & -0.07 & 0.27  & 0.65  & 0.66  & 0.69      & 0.27      & -0.04      \\
$TWN_{IJC,t}$  & 0.39  & 0.23 & -0.65 & 2.20  & -0.12 & -0.05 & 0.45  & 0.66  & 0.66  & 0.42      & 0.13      & -0.03      \\
$TWN_{FOMC,t}$ & 0.14  & 0.18 & 0.89  & 3.54  & -0.14 & -0.09 & 0.13  & 0.48  & 0.66  & 0.30      & 0.15      & 0.05       \\
$TWN_{GDP,t}$  & 0.13  & 0.11 & 1.04  & 6.54  & -0.13 & -0.05 & 0.12  & 0.31  & 0.66  & 0.24      & 0.09      & 0.07       \\
$TWN_{CPI,t}$  & 0.17  & 0.10 & 1.10  & 7.96  & -0.14 & 0.01  & 0.16  & 0.31  & 0.66  & 0.18      & 0.10      & 0.02       \\
$TWN_{ISM,t}$  & 0.18  & 0.27 & 0.71  & 1.98  & -0.12 & -0.10 & 0.08  & 0.66  & 0.66  & 0.61      & 0.22      & 0.16       \\
$TWN_{SENT,t}$ & 0.21  & 0.29 & 0.56  & 1.65  & -0.11 & -0.08 & 0.06  & 0.66  & 0.67  & 0.73      & 0.24      & 0.00       \\
$TWN_{CBCC,t}$ & 0.07  & 0.20 & 1.53  & 4.72  & -0.14 & -0.11 & -0.01 & 0.54  & 0.66  & 0.33      & 0.11      & 0.00       \\
$TWN_{RSA,t}$  & 0.04  & 0.19 & 2.13  & 6.67  & -0.11 & -0.09 & -0.04 & 0.51  & 0.66  & 0.44      & 0.17      & 0.01       \\
$TWN_{DGO,t}$  & 0.22  & 0.29 & 0.49  & 1.59  & -0.12 & -0.09 & 0.11  & 0.66  & 0.66  & 0.90      & 0.58      & 0.00       \\
\midrule
\multicolumn{13}{l}{\textit{Panel C: Twitter negative sentiment measures}}   \\
$NPP_{NFP,t}$  & -2.45 & 1.38 & 0.17  & 1.80  & -4.83 & -4.50 & -2.49 & -0.15 & -0.06 & 0.87      & 0.60      & 0.13       \\
$NPP_{IJC,t}$  & -2.66 & 1.34 & 0.31  & 1.88  & -4.88 & -4.46 & -2.92 & -0.33 & -0.05 & 0.77      & 0.19      & 0.10       \\
$NPP_{FOMC,t}$ & -2.19 & 1.18 & 0.02  & 2.12  & -4.93 & -4.20 & -2.20 & -0.20 & -0.06 & 0.74      & 0.33      & 0.05       \\
$NPP_{GDP,t}$  & 0.16  & 0.25 & 0.78  & 2.05  & -0.11 & -0.09 & 0.06  & 0.62  & 0.65  & 0.92      & 0.69      & 0.25       \\
$NPP_{CPI,t}$  & -1.92 & 1.28 & -0.25 & 1.86  & -4.75 & -4.04 & -1.84 & -0.16 & -0.06 & 0.66      & 0.20      & 0.10       \\
$NPP_{ISM,t}$  & 0.14  & 0.26 & 0.89  & 2.14  & -0.12 & -0.10 & 0.02  & 0.61  & 0.64  & 0.96      & 0.85      & 0.49       \\
$NPP_{SENT,t}$ & -2.06 & 1.43 & -0.21 & 1.69  & -4.88 & -4.34 & -2.00 & -0.11 & -0.05 & 0.89      & 0.53      & 0.05       \\
$NPP_{CBCC,t}$ & 0.11  & 0.26 & 1.00  & 2.52  & -0.15 & -0.13 & 0.03  & 0.61  & 0.64  & 0.93      & 0.71      & 0.13       \\
$NPP_{RSA,t}$  & 0.11  & 0.25 & 1.06  & 2.65  & -0.14 & -0.11 & -0.01 & 0.62  & 0.64  & 0.94      & 0.74      & 0.28       \\
$NPP_{DGO,t}$  & 0.15  & 0.24 & 0.77  & 2.20  & -0.12 & -0.10 & 0.06  & 0.62  & 0.64  & 0.92      & 0.71      & 0.10       \\
\midrule
\multicolumn{13}{l}{\textit{Panel D: News articles negative sentiment measures}}   \\
$NPN_{NFP,t}$  & -0.76 & 0.97 & -1.34 & 3.62  & -3.87 & -2.78 & -0.22 & -0.04 & -0.03 & 0.88      & 0.64      & 0.31       \\
$NPN_{IJC,t}$  & -0.47 & 0.68 & -2.76 & 11.20 & -4.46 & -2.06 & -0.18 & -0.04 & -0.03 & 0.71      & 0.13      & 0.02       \\
$NPN_{FOMC,t}$ & -1.12 & 1.09 & -0.90 & 2.70  & -4.15 & -3.31 & -0.71 & -0.05 & -0.03 & 0.73      & 0.32      & 0.13       \\
$NPN_{GDP,t}$  & 0.37  & 0.27 & -0.35 & 1.50  & -0.09 & -0.05 & 0.47  & 0.66  & 0.66  & 0.90      & 0.64      & 0.22       \\
$NPN_{CPI,t}$  & -0.92 & 0.98 & -1.23 & 3.51  & -4.17 & -3.00 & -0.56 & -0.05 & -0.03 & 0.64      & 0.19      & 0.02       \\
$NPN_{ISM,t}$  & 0.44  & 0.24 & -0.80 & 2.00  & -0.09 & -0.02 & 0.56  & 0.66  & 0.66  & 0.96      & 0.83      & 0.45       \\
$NPN_{SENT,t}$ & -0.82 & 0.96 & -1.34 & 3.81  & -3.91 & -3.07 & -0.37 & -0.04 & -0.03 & 0.90      & 0.57      & 0.13       \\
$NPN_{CBCC,t}$ & 0.44  & 0.24 & -0.81 & 2.19  & -0.11 & -0.06 & 0.54  & 0.66  & 0.66  & 0.93      & 0.71      & 0.15       \\
$NPN_{RSA,t}$  & 0.39  & 0.25 & -0.50 & 1.76  & -0.12 & -0.04 & 0.49  & 0.65  & 0.66  & 0.93      & 0.68      & 0.11       \\
$NPN_{DGO,t}$  & 0.45  & 0.23 & -0.83 & 2.35  & -0.09 & 0.00  & 0.56  & 0.66  & 0.66  & 0.92      & 0.70      & 0.10       \\
\bottomrule
\end{tabular}

\fnote{Notes: The table reports descriptive statistics (average, standard deviation, skewness, kurtosis, minimum, 5th, 50th, and 95th percentiles, maximum, and $\rho(.)$, the autocorrelation coefficient of a given order) of sentiment measures related to $10$ specific macroeconomic news announcements. Panels A-D report on positive sentiment from Twitter, positive sentiment of newspapers, negative sentiment from Twitter, and finally, negative sentiment in news articles, in this order. Note that all included sentiment indicators were constructed via the FinBERT model (see Section \ref{subsubsec:sent}). They were then aggregated over multiple keywords related to each macroeconomic news announcement (listed in Table \ref{Tbl_event_small}) and subsequently log-transformed.}

\end{table}


\clearpage
\begin{table}[!ht]

\centering
\setlength{\tabcolsep}{4pt} 
\renewcommand{\arraystretch}{1.1} 
\small

\caption{Pairwise analysis of the performance of competing models with one-sided Wilcoxon signed-rank test
}
\label{W_5w_mse_H1_all}
\small{
\begin{tabular}{l|rrr|rrrr|rrrr}
\toprule

  & \multicolumn{3}{c}{Benchmark} & \multicolumn{4}{|c}{Attention} & \multicolumn{4}{|c}{Sentiment}     \\
  & \multicolumn{1}{c}{HAR} & \multicolumn{1}{c}{CSR} & \multicolumn{1}{c}{HAR-M} & \multicolumn{1}{|c}{HAR-A} & \multicolumn{1}{c}{CSR-A} & \multicolumn{1}{c}{ALA-A} & \multicolumn{1}{c}{RF-A} & \multicolumn{1}{|c}{HAR-S} & \multicolumn{1}{c}{CSR-S} & \multicolumn{1}{c}{ALA-S} & \multicolumn{1}{c}{RF-S} \\

\midrule
HAR  &               & \bftab 0.01  & 1.00  & 1.00         & \bftab 0.00   & \bftab 0.00   & \bftab 0.00   & \bftab 0.00   & \bftab 0.00 & \bftab 0.00   & 1.00          \\
CSR  & 1.00          &              & 1.00  & 1.00         & \bftab 0.00   & \bftab 0.00   & \bftab 0.00   & \bftab 0.00   & \bftab 0.00 & \bftab 0.00   & 1.00          \\
HAR-M  & \bftab 0.00   & \bftab 0.00  &       & \bftab 0.02  & \bftab 0.00   & \bftab 0.00   & \bftab 0.00   & \bftab 0.00   & \bftab 0.00 & \bftab 0.00   & \bftab 0.00   \\
\midrule
HAR-A  & \bftab 0.00   & \bftab 0.00  & 0.98  &              & \bftab 0.00   & \bftab 0.00   & \bftab 0.00   & \bftab 0.00   & \bftab 0.00 & \bftab 0.00   & \bftab 0.00   \\
CSR-A  & 1.00          & 1.00         & 1.00  & 1.00         &               & 1.00          & 1.00          & 0.73          & \bftab 0.00 & \bftab 0.00   & 1.00          \\
ALA-A  & 1.00          & 1.00         & 1.00  & 1.00         & \bftab 0.00   &               & 1.00          & \bftab 0.00   & \bftab 0.00 & \bftab 0.00   & 1.00          \\
RF-A  & 1.00          & 1.00         & 1.00  & 1.00         & \bftab 0.00   & 1.00          &               & \bftab 0.00   & \bftab 0.00 & \bftab 0.00   & 1.00          \\
\midrule
HAR-S  & 1.00          & 1.00         & 1.00  & 1.00         & 1.00          & 1.00          & 1.00          &               & \bftab 0.00 & \bftab 0.00   & 1.00          \\
CSR-S  & 1.00          & 1.00         & 1.00  & 1.00         & 1.00          & 1.00          & 1.00          & 1.00          &             & 1.00          & 1.00          \\
ALA-S  & 1.00          & 1.00         & 1.00  & 1.00         & 1.00          & 1.00          & 1.00          & 1.00          & \bftab 0.00 &               & 1.00          \\
RF-S  & 0.59          & 0.05         & 1.00  & 1.00         & \bftab 0.00   & \bftab 0.00   & \bftab 0.00   & \bftab 0.00   & \bftab 0.00 & \bftab 0.00   &               \\  
\bottomrule
\end{tabular}
}

\fnote{Notes: The table presents p-values for the one-sided Wilcoxon signed-rank test after applying Bonferroni-Holm correction. The test was performed on pairs of models, with an alternative hypothesis that the MSE (over $404$ stocks) of the column model is significantly lower than that of the selected row model. Each column is a pairwise comparison of the column model to the rest of the competing models. All p-values $< 0.05$ are indicated in bold.} 

\end{table}

\clearpage
\section{Supplementary materials}

Supplementary material related to this article summarizes the findings of a battery of robustness checks described in section \ref{sec:robust} of the paper.

\begin{table}[ht]

\centering
\setlength{\tabcolsep}{3.5pt} 

\caption{Robustness checks: Proportion of cases the column model out-performed the benchmark HAR model.}
\label{Suppl_1}
\begin{center}
\footnotesize{
\begin{tabular}{l|rrr|rrrr|rrrr}
\toprule

                          & \multicolumn{3}{c}{Benchmark} & \multicolumn{4}{|c}{Attention} & \multicolumn{4}{|c}{Sentiment}     \\
Model                 & \multicolumn{1}{c}{HAR} & \multicolumn{1}{c}{CSR} & \multicolumn{1}{c}{HAR-M} & \multicolumn{1}{|c}{HAR-A} & \multicolumn{1}{c}{CSR-A} & \multicolumn{1}{c}{ALA-A} & \multicolumn{1}{c}{RF-A} & \multicolumn{1}{|c}{HAR-S} & \multicolumn{1}{c}{CSR-S} & \multicolumn{1}{c}{ALA-S} & \multicolumn{1}{c}{RF-S} \\
\midrule
\multicolumn{12}{l}{Panel A: Main results (one day-ahead, MSE)}                                                                                                                                                                                          \\
             &                         & 55.45                   & 16.34                     & 35.40                     & 90.59                     & 70.79                     & 70.54                    & 98.51                     & 98.76                     & 94.31                     & 49.01                    \\
\midrule
\multicolumn{12}{l}{Panel B: Alternative volatility estimators (one day-ahead, MSE)}                                                                                                                                                                                                                                                 \\
5-min, summed          &                         & 47.52                   & 23.51                     & 24.50                     & 82.92                     & 60.89                     & 75.00                    & 97.77                     & 94.55                     & 92.82                     & 56.68                    \\
1-min, weighted           &                         & 81.93                   & 18.56                     & 38.61                     & 86.14                     & 72.77                     & 73.27                    & 98.27                     & 99.26                     & 96.53                     & 48.51                    \\
1-min, summed           &                         & 65.35                   & 25.99                     & 30.69                     & 80.69                     & 58.42                     & 76.49                    & 99.01                     & 92.82                     & 92.57                     & 56.44                    \\
\midrule
\multicolumn{12}{l}{Panel C: Different sentiment analysis techniques (one day-ahead, MSE)}                                                                                                                                                                                                                                           \\
EmoLex       &                         & 55.45                   & 16.34                     & 35.40                     & 90.59                     & 70.79                     & 70.54                    & 77.97                     & 95.30                     & 89.85                     & 39.11                    \\
VADER        &                         & 55.45                   & 16.34                     & 35.40                     & 90.59                     & 70.79                     & 70.54                    & 30.69                     & 94.55                     & 87.87                     & 58.17                    \\
\midrule
\multicolumn{12}{l}{Panel D: Alternative loss functions (one day-ahead)}                                                                                                                                                                                                                                                             \\
QLIKE        &                         & 8.91                    & 23.02                     & 80.45                     & 66.83                     & 55.20                     & 5.45                     & 52.48                     & 62.87                     & 17.33                     & 2.23                     \\
MAE          &                         & 27.97                   & 2.48                      & 39.60                     & 88.86                     & 60.15                     & 73.51                    & 95.30                     & 98.76                     & 83.42                     & 59.90                    \\
MAPE         &                         & 10.40                   & 5.94                      & 58.17                     & 66.34                     & 26.73                     & 97.03                    & 71.04                     & 90.35                     & 12.87                     & 96.53                    \\
\midrule
\multicolumn{12}{l}{Panel E: Multiple-day-ahead forecasting (MSE)}                                                                                                                                                                                                                                                        \\
5-day        &                         & 92.82                   & 43.32                     & 83.66                     & 96.53                     & 70.79                     & 94.55                    & 99.01                     & 99.26                     & 99.01                     & 92.33                    \\
22-day       &                         & 92.08                   & 48.76                     & 4.95                      & 94.55                     & 1.73                      & 22.03                    & 55.69                     & 88.12                     & 16.58                     & 25.74                   \\
\bottomrule
\end{tabular}
}
\end{center}

\fnote{Notes: The table presents the out-of-sample performances of 11 competing models under various alternative specifications. The first three columns report the results for three benchmark models, the standard HAR model, the superbenchmark CSR HAR model, and the HAR-M, the standard HAR model expanded with interactions of $10$ macroeconomic dummy variables with daily realized volatility. The next four models all fall under the description "attention models", although the first model HAR-A only includes general attention, while the next three models (ALA-A, CSR-A, and RF-A) also contain measures of individual, macro-event specific attention measures, only differing in the underlying forecasting methodology. The final four columns cover our four sentiment models, again, with the first model, HAR-S, focused only on general sentiment and the remaining three models also on individual sentiment effects. All panels report proportions of cases the column model outperformed the benchmark HAR model calculated across $404$ stocks. Panel A contains the main results of one day-ahead predictions under the MSE loss function, with volatility estimator constructed at a 5-min frequency, with a weighted sum of overnight and intraday variation following \citet{Hansen2005}, and with sentiment measures constructed via the FinBERT model. Panel B covers results for alternative volatility estimators from either 1- or 5-minute frequencies and either following the weighted approach of \citet{Hansen2005} or the simpler approach of summing intraday and overnight variation of \citet{blair2010forecasting}. Results presented in Panel C differ in the applied sentiment analysis technique. In Panel D, we cover the main results evaluated with three alternative loss functions (QLIKE, MAE, and MAPE). Finally, Panel E reports the results of multiple-day-ahead predictions under the MSE loss function.}

\end{table}

\begin{table}[ht]

\centering
\setlength{\tabcolsep}{3.5pt} 

\caption{Robustness checks: Average forecast improvement of the column model compared to the benchmark HAR model.}
\label{Suppl_2}
\begin{center}
\footnotesize{
\begin{tabular}{l|rrr|rrrr|rrrr}
\toprule

                          & \multicolumn{3}{c}{Benchmark} & \multicolumn{4}{|c}{Attention} & \multicolumn{4}{|c}{Sentiment}     \\
Model                 & \multicolumn{1}{c}{HAR} & \multicolumn{1}{c}{CSR} & \multicolumn{1}{c}{HAR-M} & \multicolumn{1}{|c}{HAR-A} & \multicolumn{1}{c}{CSR-A} & \multicolumn{1}{c}{ALA-A} & \multicolumn{1}{c}{RF-A} & \multicolumn{1}{|c}{HAR-S} & \multicolumn{1}{c}{CSR-S} & \multicolumn{1}{c}{ALA-S} & \multicolumn{1}{c}{RF-S} \\
\midrule
\multicolumn{12}{l}{Panel A: Main results (one day-ahead, MSE)}                                                                                                                                                                                          \\
                &  & -0.14 & -9.38 & -5.94  & 8.12  & 5.13   & 4.61   & 8.62  & 12.74 & 11.79  & -2.15 \\
\midrule
\multicolumn{12}{l}{Panel B: Alternative volatility estimators (one day-ahead, MSE)}                    \\
5-min, summed   &  & -1.96 & -9.12 & -13.40 & 8.08  & 3.41   & 9.14   & 10.67 & 13.41 & 14.81  & 2.63  \\
1-min, weighted &  & 5.03  & -6.89 & -6.79  & 7.10  & 5.64   & 5.24   & 9.59  & 12.02 & 12.05  & -1.25 \\
1-min, summed   &  & 2.73  & -7.29 & -11.33 & 7.49  & 2.57   & 8.74   & 10.38 & 12.16 & 13.35  & 3.20  \\
\midrule
\multicolumn{12}{l}{Panel C: Different sentiment analysis techniques (one day-ahead, MSE)}               \\
EmoLex          &  & -0.14 & -9.38 & -5.94  & 8.12  & 5.13   & 4.61   & 2.42  & 8.74  & 10.02  & -4.78 \\
VADER           &  & -0.14 & -9.38 & -5.94  & 8.12  & 5.13   & 4.61   & -1.94 & 9.21  & 10.05  & 0.48  \\
\midrule
\multicolumn{12}{l}{Panel D: Alternative loss functions (one day-ahead)}                                \\
QLIKE           &  & -2.38 & -0.96 & 1.04   & 0.39  & 0.34   & -4.42  & 0.02  & 0.34  & -1.09  & -6.38 \\
MAE             &  & -0.71 & -2.01 & -0.47  & 1.65  & 0.55   & 1.47   & 1.59  & 2.56  & 1.63   & 0.41  \\
MAPE            &  & -1.67 & -0.87 & 0.27   & 0.39  & -0.83  & 5.24   & 0.39  & 1.03  & -1.37  & 5.69  \\
\midrule
\multicolumn{12}{l}{Panel E: Multiple-day-ahead forecasting (MSE)}                                                                                                                                                                                                                                                        \\
5-day           &  & 18.08 & -1.63 & 44.02  & 30.87 & 10.57  & 31.91  & 29.41 & 35.45 & 42.32  & 27.61 \\
22-day          &  & 11.54 & 0.35  & -49.67 & 15.16 & -64.87 & -12.91 & 1.86  & 11.92 & -15.14 & -5.03  \\
\bottomrule
\end{tabular}
}
\end{center}

\fnote{Notes: The table presents the out-of-sample performances of 11 competing models under various alternative specifications. The first three columns report the results for three benchmark models, the standard HAR model, the superbenchmark CSR HAR model, and the HAR-M, the standard HAR model expanded with interactions of $10$ macroeconomic dummy variables with daily realized volatility. The next four models all fall under the description "attention models", although the first model HAR-A only includes general attention, while the next three models (ALA-A, CSR-A, and RF-A) also contain measures of individual, macro-event specific attention measures, only differing in the underlying forecasting methodology. The final four columns cover our four sentiment models, again, with the first model, HAR-S, focused only on general sentiment and the remaining three models also on individual sentiment effects. All panels report an average forecast improvement of the column model compared to the benchmark HAR model calculated across $404$ stocks. Panel A contains the main results of one day-ahead predictions under the MSE loss function, with volatility estimator constructed at a 5-min frequency, with a weighted sum of overnight and intraday variation following \citet{Hansen2005}, and with sentiment measures constructed via the FinBERT model. Panel B covers results for alternative volatility estimators from either 1- or 5-minute frequencies and either following the weighted approach of \citet{Hansen2005} or the simpler approach of summing intraday and overnight variation of \citet{blair2010forecasting}. Results presented in Panel C differ in the applied sentiment analysis technique. In Panel D, we cover the main results evaluated with three alternative loss functions (QLIKE, MAE, and MAPE). Finally, Panel E reports the results of multiple-day-ahead predictions under the MSE loss function.}

\end{table}





\end{document}